\documentclass[11pt]{article}
\usepackage[margin=1in]{geometry}
\usepackage[T1]{fontenc}
\usepackage{lmodern}
\usepackage{amsmath,amssymb,bm}
\usepackage{graphicx}
\usepackage{booktabs}
\usepackage{xcolor}
\usepackage{float}
\usepackage{authblk}
\usepackage[numbers,sort&compress,merge,round]{natbib}
\usepackage[colorlinks=true, citecolor=blue, linkcolor=blue, urlcolor=blue]{hyperref}

\newcommand{\doi}[1]{}

\title{Correlation-induced suppression of Granger causality and Gaussian transfer entropy}

\author{Louis Cattafesta\thanks{To whom correspondence should be addressed. E-mail: \href{mailto:lcattafestaiii@illinoistech.edu}{lcattafestaiii@illinoistech.edu}}}
\author{Scott T.~M.~Dawson}
\author{Ankit Srivastava}
\affil{Department of Mechanical, Materials, and Aerospace Engineering, Illinois Institute of
Technology, Chicago, IL, 60616 USA}
\date{}

\begin{document}

\maketitle

\begin{abstract}
In complex dynamical systems, Granger causality and transfer entropy are widely used to infer directional interactions, but they quantify a source's unique predictive contribution after conditioning and can therefore be substantially attenuated by correlation with the retained variables. Here, using our previously developed Linear Operator Causality Analysis (LOCA) framework, which measures causal response from the system propagator, we derive a relationship between predictive and operator-based causality. For a two-dimensional linear system, the fractional suppression of Granger causality equals the squared correlation between source and receiver. We generalize this result to high-dimensional systems using the source covariance remaining after conditioning and show that additional conditioning can only increase the suppression. The same covariance ordering yields a corresponding upper bound for Gaussian transfer entropy. A linearized Burgers advection chain exhibits 36.6\% suppression at the convective travel-time peak, and exact nonlinear acoustic Burgers solutions show that the fitted-surrogate penalty is governed by correlation structure rather than nonlinearity and remains robust to sensor noise after operator debiasing. Applied to resting-state fMRI from the Human Connectome Project (80 brain regions, 8 functional networks, 10 subjects), the single-lag Granger causality measure falls 39.4\% below the operator response when conditioning only on the receiver and 73.0\% when conditioning on the receiver and all other retained variables. The framework provides a quantitative bridge between predictive and operator-based causal measures in high-dimensional dynamical systems.
\end{abstract}


\section*{Significance statement}
Directional influence can mean different things: a variable may drive a system's response yet contribute little unique predictive information once correlated variables are included. This distinction can cause predictive-causality measures to systematically understate operator-based response measures in high-dimensional systems. We show how to quantify this gap for linear dynamical models, and show that additional conditioning can only widen it. In resting-state brain imaging, conditional Granger causality falls on average 73\% below the operator response, a gap determined by how strongly brain regions co-fluctuate. The result provides a common framework for Granger causality, Gaussian transfer entropy, and operator-based causality, clarifying when they agree, why they differ, and how those differences should be interpreted across neuroscience, physics, and engineering applications.

Complex dynamical systems often exhibit intricate interactions among variables, where distinguishing causal influence from mere correlation is critical for scientific understanding and design. Granger causality \cite{granger1969} and methods based on information theory \cite{schreiber2000te,barnett2009granger} are widely used to infer directional interactions from time-series data. However, when variables are correlated, their predictive-causality estimates can differ substantially from intervention- or response-based measures \cite{stokes2017problems,barnett2018solved}. This distinction has motivated operator-theoretic and interventional alternatives, notably the Dynamical Causal Effects (DCE) framework \cite{smirnov2014dce,smirnov2015granger,smirnov2022generative}, which grounds causal inference in the system's propagator rather than statistical association. Here, building on the LOCA framework for linearized fluid flows \cite{srivastava2024loca}, itself a specific DCE, we derive the closed-form relationship between LOCA and Granger causality, an explicit penalty identity that quantifies the gap between the operator-based causal response and the source's unique predictive contribution.

\section*{Operator theoretic measure of causality}

For a continuous-time linear dynamical system $\dot{\mathbf{x}}(t) = \mathbf{A}_c\mathbf{x}(t)$, the causal influence of mode $j$ on mode $i$ at time lag $\tau$ is given by the matrix exponential element $[\mathrm{e}^{\mathbf{A}_c\tau}]_{ij}$ \cite{srivastava2024loca}. See Appendix~\ref{si:continuous} for details. Here, we develop the discrete-time formulation required for data-driven analysis.

In many practical applications, system data is collected at discrete time intervals, and the dynamics are subject to stochastic forcing. We model such a system with a discrete-time linear stochastic process:
\begin{equation}
\mathbf{x}_{k+1} = {\mathbf{A}}\mathbf{x}_k + \boldsymbol{\epsilon}_k,
\label{eq:discrete_system}
\end{equation}
where $\mathbf{x}_k \in \mathbb{R}^n$ is the state vector at time step $k$, ${\mathbf{A}} \in \mathbb{R}^{n \times n}$ is the state transition matrix (propagator), and $\boldsymbol{\epsilon}_k \in \mathbb{R}^n$ is zero-mean process noise with $\mathbb{E}[\boldsymbol{\epsilon}_k \boldsymbol{\epsilon}_j^\intercal] = \mathbf{Q}\delta_{kj}$. 

Since the noise is independent of the initial state $\mathbf{x}_0$, the causal influence of state variable $j$ on variable $i$ after $k$ steps is:
\begin{equation}
\frac{\partial x_{i,k}}{\partial x_{j,0}} = [{\mathbf{A}}^k]_{ij}.
\end{equation}
Thus, the matrix power ${\mathbf{A}}^k$ serves as the \textbf{causality matrix}. For the rest of the paper, we normalize this by the signal-to-noise ratio (SNR) to define the operator-theoretic %
causality measure:
\begin{equation}\label{eq:OpF}
F_{j\to i}^{op}=\frac{[{\mathbf{A}}]_{ij}^2\, \sigma_{x_j}^2}{\sigma_{\epsilon_i}^2},
\end{equation}
where $\sigma_{x_j}^2$ is the variance of $x_j$ and $\sigma_{\epsilon_i}^2$ is the variance of the noise component $\epsilon_i$. This definition evaluates the propagator at a single lag ($k=1$).
The notation $F^{op}$ denotes a dimensionless causality magnitude, not the finite-sample regression $F$-statistic used to test the statistical significance of Granger causality. $F^{op}$ contains no sample-size or degrees-of-freedom factor. 
$F^{op}$ asks: when the source fluctuates by a typical amount, how large is the receiver's next-step response compared with its noise floor? We restrict to off-diagonal terms ($i \neq j$) throughout, consistent with the Granger causality convention.

For the general case where we wish to quantify the causal influence of a group of source variables $\mathbf{x}_{s}$ on a group of receiver variables $\mathbf{x}_{r}$, $A_{rs}$ denotes the block of $A$ with rows in $r$ and columns in $s$, so $A_{rs}\,\delta\mathbf{x}_s$ is the one-step receiver response to a source perturbation. We average the squared causal response $\|A_{rs}\,\delta\mathbf{x}_s\|^2$ over perturbations drawn from a specified source distribution $\delta\mathbf{x}_s \sim (\mathbf{0}, \Sigma_s)$ and normalize by the noise energy (see Appendix~\ref{si:group_loca} for details):
\begin{equation}
    F^{op}_{s \to r} = \frac{\text{tr}(A_{rs}\,\Sigma_s\,A_{rs}^T)}{\text{tr}(\Sigma_{\epsilon_r})}.
    \label{eq:Fop_group}
\end{equation}
This is a SNR for the causal pathway $A_{rs}$: the expected squared causal response under the source's variability, normalized by the noise floor. For a stationary stochastic system, $\Sigma_s$ is the stationary source covariance. For a transient data ensemble, it is the sample covariance of the pooled source perturbations, as in the acoustic example below.

\section*{Granger causality for a two-state system}

We now analyze a two-dimensional discrete-time linear system whose causal properties have been studied in \cite{barnett2011behaviour,barnett2014mvgc,barnett2015granger}:
\begin{equation}
\label{eq:2d_system}
    \begin{bmatrix} x_{k+1} \\ y_{k+1}\end{bmatrix}
    = \begin{bmatrix}
 a & c \\
 0 & b
    \end{bmatrix} \begin{bmatrix} x_{k} \\ y_{k}\end{bmatrix}
    +\begin{bmatrix} \epsilon_{x,k} \\ \epsilon_{y,k}
    \end{bmatrix}.
\end{equation}
We assume that the system is stable, with $|a|, |b| < 1$. The only nontrivial causal connection is the effect of $y$ on $x$, governed by $c\neq 0$. From Eq.~\ref{eq:OpF}, 
\begin{equation}
\label{eq:Fop_2d}
    F^{op}_{y\to x} = \frac{c^2\,\text{var}(y)}{\sigma_{\epsilon,x}^2}
    = \frac{c^2}{1-b^2}\frac{\sigma_{\epsilon,y}^2}{\sigma_{\epsilon,x}^2},
\end{equation}
where the second equality follows from $\text{var}(y) = \sigma_{\epsilon,y}^2/(1-b^2)$.

A closed-form expression for the corresponding Granger causality measure %
 exists \cite{barnett2014mvgc} and depends on $a$, whereas $F^{op}_{y\to x}$ does not (Appendix~\ref{si:barnett}). To see why, consider how this Granger causality measure is constructed. Granger causality compares an \textbf{unrestricted model} that includes all variables (here simply Eq.~\ref{eq:2d_system}) against a \textbf{restricted model} that drops the variable whose causal influence is being tested:
\begin{equation}
\label{eq:restricted}
     x_{k+1} = \tilde{a} x_{k}+\tilde{\epsilon}_{x,k}.
\end{equation}
The normalized variance-increment Granger causality measure, $F^{GC}$, is the relative increase in one-step prediction error:
\begin{equation}\label{eq:GCF}
     F_{y\to x}^{GC} = \frac{\tilde{\sigma}_{\epsilon,x}^2 - \sigma_{\epsilon,x}^2}{\sigma_{\epsilon,x}^2}.
\end{equation}
This normalized variance increment is distinct from the finite-sample regression $F$-statistic used for significance testing, although both are based on the same reduction in prediction error. It is also a monotonic transformation of the Granger--Geweke log variance-ratio measure $\mathcal{F}_{y\to x}=\ln(\tilde{\sigma}_{\epsilon,x}^2/\sigma_{\epsilon,x}^2) = \ln(1+F^{GC}_{y\to x})$ \cite{geweke1982measurement}. 
The key point is that $\tilde{a} \neq a$: the omitted variable $y$ is correlated with $x$, so the restricted least-squares estimate absorbs part of the omitted effect into the self-coupling coefficient (Appendix~\ref{si:least_squares}):
\begin{equation}
\label{eq:ar_biased}
    \tilde{a} = a + c \frac{\text{cov}(x,y)}{\text{var}(x)}.
\end{equation}
Dropping $y$ does not remove its influence from the restricted fit. The component of that influence that is predictable from $x$ is absorbed into the fitted self-coupling coefficient, and only the linearly unpredictable remainder is attributed to $y$. The restricted error is therefore not simply $\epsilon_{x,k}$ but includes the part of $y$'s influence uncorrelated with $x$:
\begin{equation}
    \tilde{\epsilon}_{x,k} = \epsilon_{x,k} + c\left(y_k -
    \frac{\text{cov}(x,y)}{\text{var}(x)}x_k\right).
\end{equation}
The term in parentheses is the part of \(y_k\) that cannot be predicted linearly from \(x_k\), consistent with the Frisch--Waugh--Lovell theorem \cite{frisch1933partial,lovell1963seasonal}, which states that the incremental effect of a predictor is determined by the component remaining after projecting out the other retained predictors.  
Computing the variance of the restricted error above (Appendix~\ref{si:restricted_variance}) and substituting into Eq.~\ref{eq:GCF} yields:
\begin{equation}\label{eq:GC_general}
     F_{y\to x}^{GC} = \frac{c^2}{\sigma_{\epsilon,x}^2}\left(\text{var}(y)-
     \frac{\text{cov}(x,y)^2}{\text{var}(x)}\right).
\end{equation}
Comparing with Eq.~\ref{eq:Fop_2d} and factoring out $\text{var}(y)$, we arrive at the central result:
\begin{equation}\label{eq:CG-op}
     F_{y\to x}^{GC} = F_{y\to x}^{op}\left[1-\rho^2(x,y)\right],
\end{equation}
where $\rho(x,y)$ is the Pearson correlation between $x$ and $y$. We define the \textbf{correlation penalty} $\mathcal{P} \equiv \rho^2$: the fraction of the operator response that the Granger statistic does not attribute to the source. The inverse relation $F^{op}_{y\to x} = F^{GC}_{y\to x}/(1-\rho^2)$ gives a direct correction: a Granger estimate plus the pairwise correlation recovers the suppressed LOCA metric, although the correction amplifies uncertainty in the estimated $\rho$ by $1/(1-\rho^2)$ and so degrades as $|\rho|\to 1$. Granger causality vanishes in the limit $|\rho|\to 1$, where the regression becomes singular and the source is no longer separately identifiable, and coincides with LOCA only at $\rho=0$; $F^{op}$ is thus the upper bound on $F^{GC}$. Standard Granger causality conditions on the entire past. For fully observed VAR(1) systems, the unrestricted one-lag predictor is already optimal, and restricted-model history can only shrink the restricted error, so $F^{op}$ also bounds the full-history statistic (Appendix~\ref{si:fullhistory}).

\begin{figure}[t!]
\centering
\null\hfill
(a)\includegraphics[width = 0.25\linewidth]{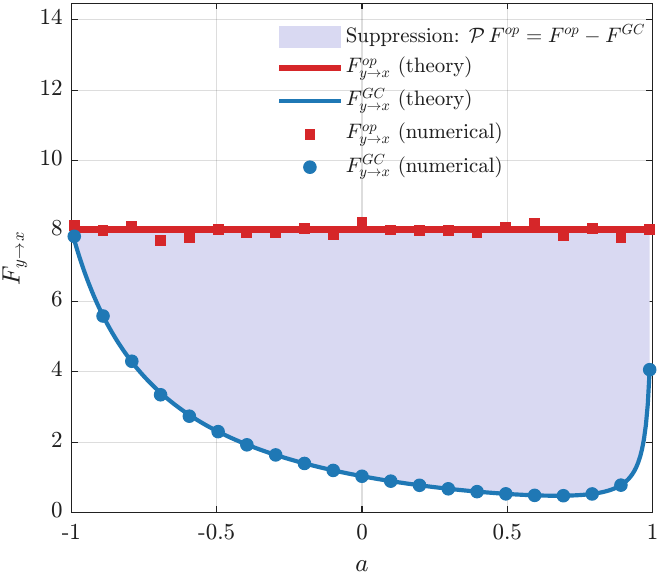}%
\hfill
(b)\includegraphics[width = 0.3\linewidth]{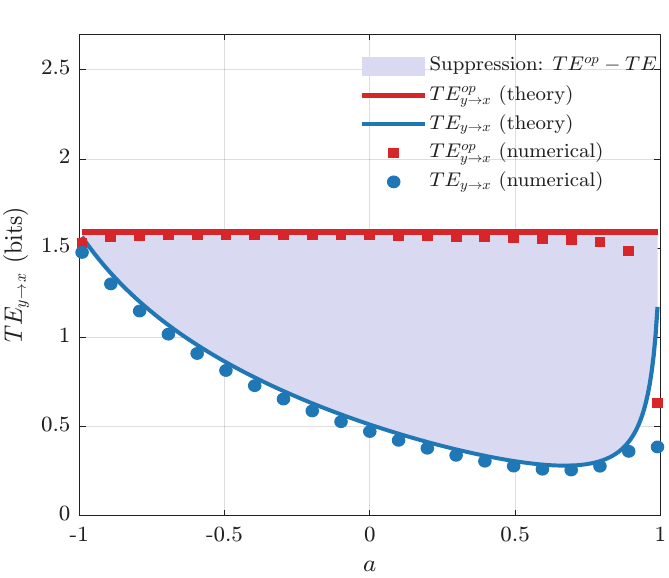}%
\hfill\null
\caption{(a) Granger versus operator-based causality for the system of \eqref{eq:2d_system} with $b = 0.99$, $c = 0.4$, varying $a$, and uncorrelated Gaussian noise, $\text{var}(\epsilon_x) = \text{var}(\epsilon_y) = 0.01$. (b) Transfer entropy for the same system. Numerical estimates use $10^7$ timesteps (first $2\times 10^4$ discarded); TE uses shuffle-corrected quantile binning (Appendix~\ref{si:te}).}
\label{fig:2Dsys}
\end{figure}

Figure \ref{fig:2Dsys}(a) illustrates this for \eqref{eq:2d_system} across a range of $a$ at fixed $b$ and $c$: $F^{op}$ is independent of $a$ while $F^{GC}$ varies through the $a$-dependent correlation between $x$ and $y$, and theory agrees closely with values from standard linear regression. 

\section*{Generalizing to N-dimensional systems}

The preceding analysis generalizes to $N$-dimensional systems via a three-way partition. Let $s$, $r$, and $o$ denote disjoint index sets for \textbf{source}, \textbf{receiver} (target), and \textbf{other} (conditioning) variables, with $s\cup r\cup o = \{1,\ldots,N\}$. Roman entrywise indices $i,j$, as in Eq.~\ref{eq:OpF} and ref.~\cite{srivastava2024loca}, denote single matrix elements. The set subscripts $s$, $r$, and $o$ denote blocks. Define the combined retained predictor $\mathbf{z}_k = [\mathbf{x}_{r,k};\,\mathbf{x}_{o,k}]$ and write the \textbf{unrestricted model}:
\begin{equation}
    \mathbf{x}_{r,k+1} = A_{rz}\mathbf{z}_k + A_{rs}\mathbf{x}_{s,k} + \boldsymbol{\epsilon}_{r,k}, \qquad A_{rz} = \begin{bmatrix}A_{rr} & A_{ro}\end{bmatrix},
\end{equation}
with error covariance $\Sigma_{\epsilon_r}$. The \textbf{restricted model} retains $\mathbf{z}_k$ but omits $\mathbf{x}_s$:
\begin{equation}
    \mathbf{x}_{r,k+1} = \tilde{A}_{rz}\,\mathbf{z}_k + \tilde{\boldsymbol{\epsilon}}_{r,k}.
\end{equation}
Least squares absorbs the source component predictable from $\mathbf{z}_k$ into the restricted coefficient (Appendix~\ref{si:ares}):
\begin{equation}
    \tilde{A}_{rz} = A_{rz} + A_{rs}\Sigma_{sz}\Sigma_z^{-1},
    \label{eq:Atilde_rz}
\end{equation}
where $\Sigma_z = \mathbb{E}[\mathbf{z}_k\mathbf{z}_k^T]$ and $\Sigma_{sz} = \mathbb{E}[\mathbf{x}_{s,k}\mathbf{z}_k^T]$. The restricted residual $\tilde{\boldsymbol{\epsilon}}_{r,k} = A_{rs}\mathbf{x}_{s,k}^* + \boldsymbol{\epsilon}_{r,k}$, where $\mathbf{x}_{s,k}^* = \mathbf{x}_{s,k} - \Sigma_{sz}\Sigma_z^{-1}\mathbf{z}_k$ is the source component orthogonal to all retained predictors. Its covariance is (Appendix~\ref{si:restricted_covariance}):
\begin{equation}
    \Sigma_{\tilde{\epsilon}_r} = A_{rs}\underbrace{\left(\Sigma_s - \Sigma_{sz}\Sigma_z^{-1}\Sigma_{zs}\right)}_{\Sigma_{s|(r,o)}}A_{rs}^T + \Sigma_{\epsilon_r},
    \label{eq:Sigma_tilde_main}
\end{equation}
where $\Sigma_{s|(r,o)}$ is the partial covariance of $\mathbf{x}_s$ given the full retained predictor set $\mathbf{z}_k$. The corresponding trace-based conditional Granger variance-increment measure is %
\begin{equation}\label{eq:Granger-N-dim-F}
    F_{s\to r|o}^{GC} = \frac{\text{tr}\!\left(A_{rs}\,\Sigma_{s|(r,o)}\,A_{rs}^T\right)}{\text{tr}(\Sigma_{\epsilon_r})}.
\end{equation}
Comparing with $F^{op}_{s\to r} = \text{tr}(A_{rs}\,\Sigma_s\,A_{rs}^T)/\text{tr}(\Sigma_{\epsilon_r})$, which uses the full unconditional source covariance $\Sigma_s$, their difference is:
\begin{equation}\label{eq:bias}
    F^{op}_{s\to r} - F^{GC}_{s\to r|o} = \frac{\text{tr}\!\left(A_{rs}\,\Sigma_{sz}\Sigma_z^{-1}\Sigma_{zs}\,A_{rs}^T\right)}{\text{tr}(\Sigma_{\epsilon_r})} \geq 0,
\end{equation}
where non-negativity follows from $\Sigma_{sz}\Sigma_z^{-1}\Sigma_{zs}$ being positive semidefinite, establishing $F^{op}_{s\to r}$ as a \emph{general upper bound} on $F^{GC}_{s\to r|o}$. The \textbf{N-dimensional correlation penalty} is:
\begin{equation}\label{eq:penalty_Ndim}
    \mathcal{P}_{s\to r|o} = \frac{F^{op}_{s\to r} - F^{GC}_{s\to r|o}}{F^{op}_{s\to r}} = \frac{\text{tr}\!\left(A_{rs}\,\Sigma_{sz}\Sigma_z^{-1}\Sigma_{zs}\,A_{rs}^T\right)}{\text{tr}\!\left(A_{rs}\,\Sigma_s\,A_{rs}^T\right)},
\end{equation}
which reduces to $\rho^2$ in the scalar two-partition case. Setting $\mathbf{x}_o = \emptyset$ recovers the two-partition result ($\Sigma_z\to\Sigma_r$, $\Sigma_{sz}\to\Sigma_{sr}$). Conditioning on additional variables can only increase the penalty, yielding the monotonic ordering:
\begin{equation}\label{eq:ordering}
    F_{s\to r|o}^{GC} \leq F_{s\to r}^{GC} \leq F_{s\to r}^{op},
\end{equation}
since $\Sigma_{s|(r,o)}\preceq\Sigma_{s|r}\preceq\Sigma_s$, where $\preceq$ denotes the positive-semidefinite ordering. Geometrically, LOCA averages the squared causal response over the full source perturbation space, while conditional Granger averages only over the subspace orthogonal to all retained predictors. The variance in the correlated subspace constitutes the penalty. Adding conditioning variables can only reduce the source covariance left unexplained by the retained predictors. With the same full-state propagator block $A_{rs}$, this can only decrease the Granger causality measure relative to the operator response. Hereafter the subscripts $2p$ and $3p$ abbreviate the two- and three-partition statistics, $F^{GC}_{2p} \equiv F^{GC}_{s\to r}$ and $F^{GC}_{3p} \equiv F^{GC}_{s\to r|o}$, with corresponding penalties $\mathcal{P}_{2p}$ and $\mathcal{P}_{3p}$.

\section*{Connection to transfer entropy}

Transfer entropy (TE) \cite{schreiber2000te} is formally connected to Granger causality: for jointly Gaussian processes the two are equivalent, with the scalar relation \cite{barnett2009granger} (using $\log_2$ throughout)
\begin{equation}\label{eq:TE_GC_relation}
    TE_{y\to x} = \tfrac{1}{2}\log_2\!\left(1 + F^{GC}_{y\to x}\right).\end{equation}
Substituting $F^{GC}_{y\to x} = F^{op}_{y\to x}(1-\rho^2)$ from Eq.~\ref{eq:CG-op}, and defining the \textbf{operator-theoretic transfer entropy} $TE^{op}_{y\to x} \equiv \tfrac{1}{2}\log_2(1 + F^{op}_{y\to x})$, we obtain
\begin{equation}\label{eq:TE_bound}
    TE_{y\to x} = \tfrac{1}{2}\log_2\!\left[1 + F^{op}_{y\to x}(1-\rho^2)\right] \;\leq\; TE^{op}_{y\to x},
\end{equation}
with equality if and only if $\rho = 0$.

The result extends to $N$ dimensions: the scalar variance ratio in Eq.~\ref{eq:TE_GC_relation} generalizes to a determinant ratio, $TE_{s\to r|o} = \tfrac{1}{2}\log_2(|\Sigma_{\tilde{\epsilon}_r}|/|\Sigma_{\epsilon_r}|)$, and inserting $\Sigma_{\tilde{\epsilon}_r}$ from above yields
\begin{equation}\label{eq:TE_Ndim}
    TE_{s\to r|o} = \tfrac{1}{2}\log_2\!\left|\mathbf{I} + \Sigma_{\epsilon_r}^{-1} A_{rs}\,\Sigma_{s|(r,o)}\,A_{rs}^T\right| \;\leq\; TE^{op}_{s\to r} \equiv \tfrac{1}{2}\log_2\!\left|\mathbf{I} + \Sigma_{\epsilon_r}^{-1} A_{rs}\,\Sigma_s\,A_{rs}^T\right|,
\end{equation}
where the inequality follows because conditioning can only reduce the source covariance, $\Sigma_{s|(r,o)}\preceq\Sigma_s$. 
Equality holds when $A_{rs}\,\Sigma_{sz}\Sigma_z^{-1}\Sigma_{zs}\,A_{rs}^T = \mathbf{0}$; $\Sigma_{sz}=\mathbf{0}$ is sufficient but not necessary (see Appendix~\ref{si:te_multivariate} for the formal matrix argument). Up to a factor $2\ln 2$, Eq.~\ref{eq:TE_Ndim} has the log-determinant form of a one-step Gaussian Granger/transfer-entropy statistic \cite{geweke1984measures}. Conventional Geweke causality uses the optimal restricted predictor over the full retained history. For a fully observed VAR(1), it is no greater than Eq.~\ref{eq:TE_Ndim}, so the operator bound carries over (Appendices~\ref{si:fullhistory} and~\ref{si:te_multivariate}).

Figure~\ref{fig:2Dsys}(b) shows the TE quantities for the same system: $TE^{op}_{y\to x}$ is independent of $a$ while $TE_{y\to x}$ varies through the correlation penalty, coinciding only at $\rho = 0$. The numerical estimates (shuffle-corrected binned TE, via Eq.~\ref{eq:TE_bound}) agree with theory except near $a \approx \pm 1$, where binned TE estimators converge slowly for highly autocorrelated processes \cite{mcallester2020formal,kraskov2004estimating,munoz2025tende}.

\section*{Finite-horizon causal propagation in a linearized Burgers chain}

As a minimal physics-based example, we use the viscous Burgers equation, $u_t+u u_x=\nu u_{xx}$, the simplest nonlinear model of the competition between convection and diffusion \cite{burgers1948mathematical}. Writing $u=U_0+q$ for a small perturbation $q$ about a uniform flow $U_0$ and dropping the quadratic term gives stochastically forced linear advection--diffusion, $q_t+U_0q_x=\nu q_{xx}+\eta$ (Appendix~\ref{si:burgers_derivation}). In the convection-dominated limit, first-order upwinding gives the open downstream chain
\begin{equation}
    x_{m,k+1}=a x_{m,k}+c x_{m-1,k}+\epsilon_{m,k},\qquad m=1,\ldots,N,
\end{equation}
or $\mathbf{x}_{k+1}=A\mathbf{x}_k+\boldsymbol{\epsilon}_k$ with $A=aI+cS$, where $S$ shifts disturbances downstream. Each cell keeps a fraction $a=1-U_0\Delta t/\Delta x$ of its own value and receives a fraction $c=U_0\Delta t/\Delta x$ from its upstream neighbor, so a disturbance is handed down the chain at speed $U_0$. A cell $d$ steps downstream feels the largest response near the travel time $(d\Delta x)/U_0$ (Appendix~\ref{si:burgers_horizon}).

\begin{figure}[h!]
\centering
\includegraphics[width=0.95\linewidth]{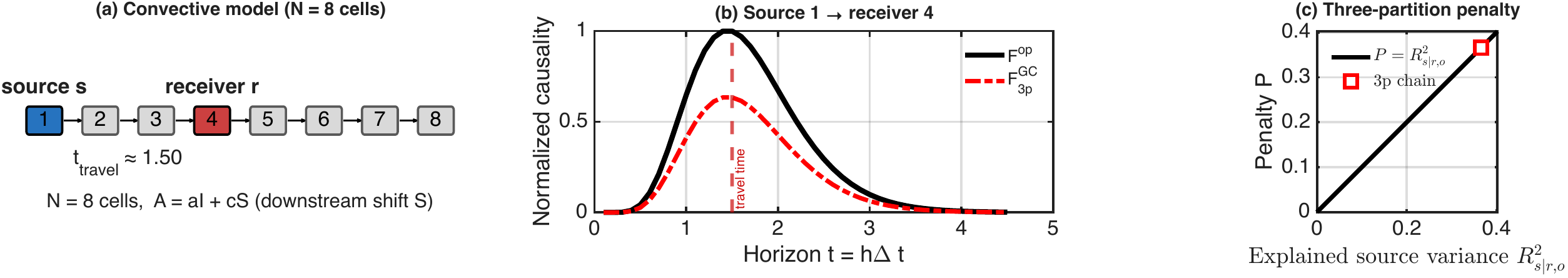}
\caption{Finite-horizon LOCA in a convective Burgers chain: (a) linearized Burgers dynamics as an $N=8$ downstream chain; (b) $F^{op}$ and $F^{GC}_{3p}$ versus horizon $h$, each normalized by the peak value of $F^{op}$ over the horizon sweep, so the $F^{op}$ curve attains 1 at its equal discrete maxima at $h=14$ and $15$, while $F^{GC}_{3p}$ is suppressed by conditioning; (c) the penalty $P=(F^{op}-F^{GC}_{3p})/F^{op}$ equals $\mathcal{P}_{s\to r|o}$, the multiple $R^2$ of the source on the retained variables.}
\label{fig:burgers}
\end{figure}

Figure~\ref{fig:burgers} uses $N=8$, $a=0.8$, $c=0.2$, $U_0=2$, $\Delta x=1$, and $\Delta t=0.1$, with source cell 1 and receiver cell 4. The discrete LOCA response has equal maxima at $h=14$ and $h=15$. The latter coincides with the travel time $3\Delta x/U_0=1.5$\,s. With upstream-dominated white noise $Q=\mathrm{diag}(10,1,\ldots,1)$, retained downstream cells contain delayed copies of the source, so the conditioning set has already seen most of what the source will do, and $F^{GC}_{3p}/F^{op}=0.634$ at the peak: a $36.6\%$ penalty, $\mathcal{P}_{s\to r|o}=0.366$, which for this scalar source equals the multiple $R^2$ of the source on the retained variables, plotted in Fig.~\ref{fig:burgers}(c). Appendix~\ref{si:burgers} gives the full derivation, $N$-sensitivity, and nonlinear Burgers validation.

\section*{An exact nonlinear test: finite-amplitude acoustic Burgers propagation}

\begin{figure}[t!]
\centering
\includegraphics[width=0.95\linewidth]{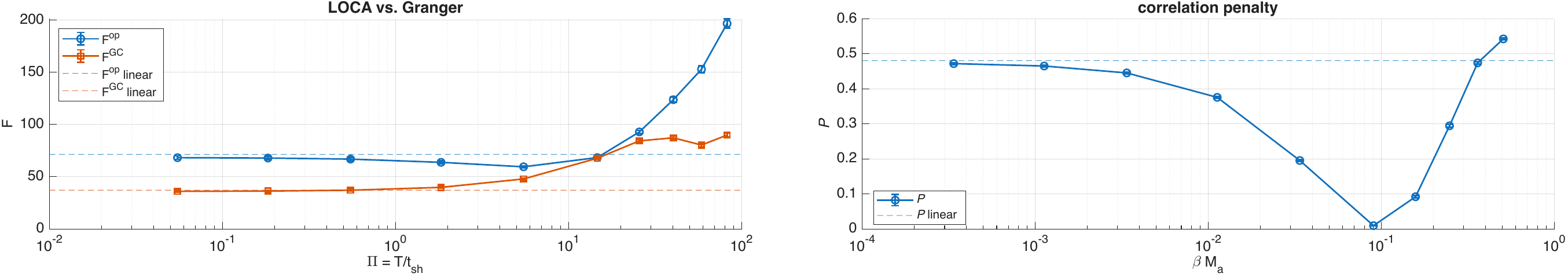}
\caption{Finite-amplitude acoustic Burgers sweep, each point averaged over 8 seeds with standard errors. Left: $F^{op}$ and $F^{GC}$ versus shock proximity $\Pi = T/t_{sh}$, with their linear asymptotes. Right: the correlation penalty $\mathcal{P}$ (axis label $P$) versus the nonlinearity parameter $\beta M_a$, showing the non-monotone plateau--collapse--recovery structure. The nonlinearity residual $\mathcal{R}$ and the machine-precision identity check are shown in Appendix~\ref{si:acoustic}.}
\label{fig:acoustic}
\end{figure}

The examples above rely on linear or linearized propagators. To test whether the correlation-penalty structure survives genuine nonlinearity and measurement noise, we generate data from a nonlinear continuum system: a finite-amplitude plane acoustic wave with thermoviscous absorption (viscous and heat-conduction damping), whose particle velocity obeys the viscous Burgers equation of nonlinear acoustics in the frame moving at the small-signal sound speed \cite{burgers1948mathematical,hamilton1998nonlinear}. Every ensemble member is an exact Cole--Hopf solution \cite{cole1951quasilinear,hopf1950partial}. Stochasticity enters through a randomized source (amplitude and location jitter) rather than through forcing of the PDE, which would destroy the exact solution. A chain of eight probes is sampled stroboscopically at the linear inter-probe transit time, and the single-lag propagator and covariances are identified from the pooled ensemble by least squares (exact DMD \cite{schmid2010dynamic,tu2014dynamic}). The amplitude sweep spans nonlinearity parameters $\beta M_a$ from $3\times 10^{-4}$ to $0.51$, where $M_a=\max|u_0|/c_0$ is the peak acoustic Mach number and $\beta$ the coefficient of nonlinearity (Appendix~\ref{si:acoustic_formulation}), equivalently shock-proximity ratios $\Pi = T/t_{sh}$ from $0.06$ (linear acoustics) to $83$ (mature $N$-wave), where $t_{sh}$ is the shock-formation time. The linear reference operator $A_L$, identified from linearized (advection--diffusion) data on the identical probe array and ensemble design, serves as the linear baseline. The nonlinearity enters through the data and is quantified by the operator residual $\mathcal{R} = \|A_{\mathrm{clean}}-A_L\|_F/\|A_L\|_F$.

Three results emerge (Fig.~\ref{fig:acoustic}; Appendix~\ref{si:acoustic}). First, for the fitted surrogate the three-partition identity $F^{GC}_{s\to r|o} = F^{op}_{s\to r}(1-\mathcal{P}_{s\to r|o})$, with $\mathcal{P}_{s\to r|o}$ the penalty of Eq.~\ref{eq:penalty_Ndim}, holds at every amplitude. The identity is a property of the least-squares fit itself, so the sweep tests how its ingredients evolve with the dynamics (Appendix~\ref{si:acoustic_quantifiers}). Second, nonlinearity and penalty decouple. The residual $\mathcal{R}$ grows monotonically from $0.013$ to $0.586$ across the sweep, whereas $\mathcal{P}$ is non-monotonic: it sits at the linear near-shift plateau ($\approx 0.47$), collapses to $\approx 0.01$ near shock formation, where the front steepens until small shifts in source location displace it by more than its own thickness so that neighboring probes no longer move together, and recovers to $\approx 0.54$ once the waveform settles into a mature $N$-wave whose large-scale sawtooth shape carries all probes coherently again. The gap between operator and Granger causality is therefore governed by the correlation structure of the measurements, not by the amount of nonlinearity. Third, additive sensor noise biases the identification through an errors-in-variables mechanism (Appendix~\ref{si:acoustic_noise}, Fig.~\ref{fig:si_acoustic_noise} and Table~\ref{tab:si_acoustic_noise}): exact DMD attenuates $F^{op}$ from $124$ to $3.9$ as the noise reaches $30\%$ of signal rms, whereas debiasing that subtracts the known noise covariance from the identification \cite{dawson2016dmdnoise} recovers the operator metric through moderate noise. At $30\%$ the corrected estimate shows greater variance, though it remains far less attenuated than exact DMD. The ratio-based penalty is comparatively stable, holding at $\mathcal{P} = 0.295$ across the entire noise range for the debiased estimator, making it far more noise-robust than the absolute causality magnitudes.

\section*{Application to resting-state fMRI}

As a test beyond controlled physical models, we turn to resting-state fMRI, where Granger causality is widely used for directed connectivity \cite{seth2015granger,seth2013granger_fmri,bielczyk2018parietal}, despite ongoing debate \cite{stokes2017problems,barnett2018solved}. Our analysis isolates a distinct issue: correlation-induced suppression arising within the fitted VAR(1) representation, the lag order selected by BIC for every subject (Appendix~\ref{si:fmri_diagnostics}). We analyze HCP resting-state BOLD data \cite{vanessen2013hcp,smith2013rsfmri,roffet2024}: 80 regions \cite{desikan2006automated}, 10 subjects, TR $=0.72$ s, and 1200 samples. Regions are grouped into seven Yeo cortical networks plus a subcortical group \cite{yeo2011organization}. See Appendix~\ref{si:hcp_methods} for data and method details. For each subject, we z-score the time series and fit the full 80-variable observed VAR(1), obtaining $\mathbf{A}$, $\Sigma$, and $\Sigma_\epsilon$.

\begin{figure}[t]
\centering
\includegraphics[width=1.0\linewidth]{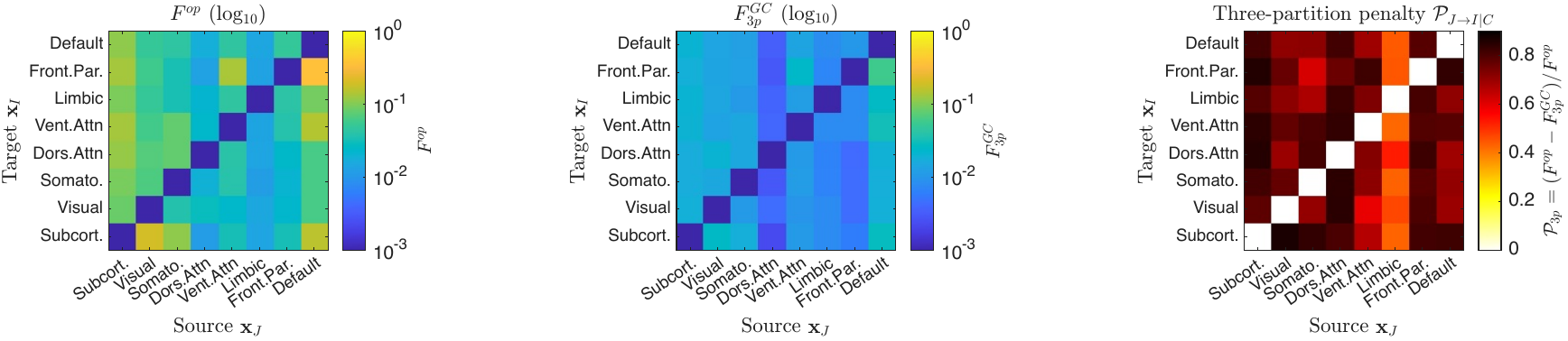}
\caption{Block-level LOCA and Granger causality on HCP resting-state fMRI; source networks in columns, receivers in rows. (a) $F^{op}$ ($\log_{10}$ scale). (b) Three-partition $F^{GC}_{3p}$, same scale. (c) Penalty $\mathcal{P}_{3p}$ (mean $73.0\%$; Limbic sources anomalously low). Labels: Subcort., Visual, Somato., Dors.Attn, Vent.Attn, Limbic, Front.Par., Default.}
\label{fig:hcp}
\end{figure}

For the block-level test, each functional network is an index set. We compute $F^{op}_{s\to r}$ and $F^{GC}_{s\to r|o}$ for all 56 directed network pairs, with $\mathbf{x}_o$ containing all regions outside the source and receiver networks. Statistics are averaged across subjects before penalty ratios are formed. Figure~\ref{fig:hcp} summarizes the network results. $F^{op}$ identifies Default Mode $\to$ Frontoparietal as the dominant link ($5.6\times$ the mean). 
Because these block measures aggregate over source channels, larger source networks can attain larger values; the Default Mode $\to$ Frontoparietal link remains dominant after normalization by source-block size (Appendix~\ref{si:hcp_block}). 
The three-partition Granger map preserves the spatial pattern but is uniformly suppressed: $\mathcal{P}_{3p}$ averages $73.0\%$, versus $39.4\%$ for two-partition conditioning, and the ordering in Eq.~\ref{eq:ordering} holds for all 56 pairs, as it must for a fully observed fitted model. The large penalties reflect the strongly shared fluctuations of resting-state BOLD signals, which leave little unique variance for the predictive score. The anomalously low Limbic penalties may reflect sparser long-range limbic connectivity. On the one-lag log-determinant (Gaussian TE) scale of Eq.~\ref{eq:TE_Ndim} the penalties average $69.7\%$ and $34.8\%$, which by Appendix~\ref{si:fullhistory} are lower bounds on the attenuation of conventional full-history Geweke causality. Note that since the empirical BOLD residuals are close to but not exactly white, this lower-bound interpretation is exact for the fitted VAR(1) model, but only approximate for the underlying BOLD process.

\section*{Discussion}

We have shown that correlation creates a systematic gap between the operator-based causal response and the conditional predictive-causality measures analyzed here, with LOCA as the operator-theoretic upper bound: in the scalar case $F^{GC}=F^{op}(1-\rho^2)$, and in $N$ dimensions the penalty is the pathway-weighted fraction of source covariance explained by the retained predictors, yielding the ordering $F^{GC}_{s\to r|o}\leq F^{GC}_{s\to r}\leq F^{op}_{s\to r}$, which holds across all fully observed examples. Granger causality credits the source only with variance that nothing else could have predicted, so fluctuations shared with the retained variables are subtracted before scoring. LOCA instead asks how strongly the receiver responds when the source fluctuates with its full covariance, shared or not. In strongly correlated systems, such as convective flows or brain networks under common drive, most source variability is predictable from elsewhere, and the predictive score keeps only the small unique remainder.

Several limitations remain. In a fully observed, correctly specified VAR(1) setting, additional lags can only further condition the restricted Granger model and therefore increase the penalty, making the present estimates conservative (formalized in Appendix~\ref{si:fullhistory}). LOCA also requires a known or estimated propagator, so finite-data uncertainty must be quantified in applications. For nonlinear systems, the identified propagator is a linear surrogate of the true dynamics: the identity and the penalty are properties of that surrogate, and the nonlinearity appears as a growing operator residual. Similarly, the transfer-entropy statements are exact for jointly Gaussian processes. For non-Gaussian data they describe the Gaussian/log-determinant surrogate rather than the full nonparametric transfer entropy. Finally, the monotonic ordering assumes the model state vector is fully observed. Reduced-state VARs introduce omitted-variable coefficient bias and residual inflation beyond the correlation penalty. These results motivate reanalysis of Granger-based studies in neuroscience, climate, and engineering, where predictive-causality estimates can substantially understate the operator-based causal response.

\subsection*{Data, Materials, and Software Availability}
The preprocessed HCP fMRI data are available on Zenodo \cite{roffet2024}. MATLAB code reproducing all figures and appendix tables is on GitHub (\url{https://github.com/lcattafesta1/pnas-loca-release}), archived on Zenodo \cite{cattafesta2026code}.

\subsection*{Acknowledgments}
A.S.\ acknowledges support from the US National Science Foundation award \#2219203. S.D.\ acknowledges support from the US Department of Energy award DE-SC0025597 and the US National Science Foundation award \#2238770. L.C.\ acknowledges support from Air Force Office of Scientific Research Grant FA9550-23-1-0269 monitored by Dr.\ Gregg Abate.

\subsection*{Author contributions}
L.C. developed the N-dimensional extension and the acoustic example. S.D. developed the theoretical framework. A.S. contributed the fMRI example and the transfer entropy connection. All authors contributed to analysis and writing.

\subsection*{Competing interests}
The authors declare no competing interests.

\appendix
\counterwithin{figure}{section}
\counterwithin{table}{section}
\numberwithin{equation}{section}

\section{Continuous-time formulation of LOCA}
\label{si:continuous}
This section records the continuous-time origin of the LOCA framework and its connection to the discrete-time model of Eq.~\eqref{eq:discrete_system} of the main text.
For a linear dynamical system $\dot{\mathbf{x}}(t) = \mathbf{A}_c\mathbf{x}(t)$, the solution is given by the matrix exponential:
\begin{equation}
\mathbf{x}(t) = \mathrm{e}^{\mathbf{A}_ct}\mathbf{x}(0).
\end{equation}
The causal influence of mode $j$ on mode $i$ at time lag $\tau$ is therefore:
\begin{equation}
\frac{\partial x_i^{\tau}}{\partial x_j^{0}} = [\mathrm{e}^{\mathbf{A}_c\tau}]_{ij}.
\end{equation}
If this quantity is zero, mode $x_j$ has no causal influence on mode $x_i$ at lag $\tau$. The matrix exponential $\mathrm{e}^{\mathbf{A}_c\tau}$ thus encapsulates all causal relationships between system modes at any future time lag. See \cite{srivastava2024loca} for a detailed discussion of the continuous-time formulation, including its physical interpretation and relationship to resolvent-based analysis.

\section{Group-to-group LOCA metric derivation}
\label{si:group_loca}
This section derives the group-to-group statistic $F^{op}_{s\to r}$, Eq.~\eqref{eq:Fop_group} of the main text, by averaging the squared causal response over source perturbations.
For the general case where we wish to quantify the causal influence of a group of source variables $\mathbf{x}_{s}$ on a group of receiver variables $\mathbf{x}_{r}$, the relevant quantity is the propagator block $A_{rs}$. The causal response in $\mathbf{x}_r$ due to a perturbation $\delta\mathbf{x}_s$, with all other variables held fixed, is:
\begin{equation}
\delta\mathbf{x}_{r,k+1} = A_{rs}\,\delta\mathbf{x}_s,
\end{equation}
with squared magnitude $\|\delta\mathbf{x}_{r,k+1}\|^2 = \delta\mathbf{x}_s^T A_{rs}^T A_{rs}\,\delta\mathbf{x}_s$. This response energy depends on the particular source perturbation $\delta\mathbf{x}_s$, so we define a single pathway-strength measure by averaging over source perturbations with covariance $\Sigma_s$. For a stationary stochastic system, $\Sigma_s$ is the stationary source covariance. For a transient data ensemble, $\Sigma_s$ is the sample covariance of the source perturbation vectors after pooling the ensemble members and snapshots used to fit or evaluate the operator:
\begin{equation}
\mathbb{E}\left[\|\delta\mathbf{x}_{r,k+1}\|^2\right] = \mathbb{E}\left[\delta\mathbf{x}_s^T A_{rs}^T A_{rs}\,\delta\mathbf{x}_s\right].
\end{equation}
The right-hand side is the expectation of a scalar quadratic form $\mathbf{z}^T M \mathbf{z}$ with $\mathbf{z} \sim (\mathbf{0},\Sigma_s)$ and $M = A_{rs}^T A_{rs}$. Using the standard identity $\mathbb{E}[\mathbf{z}^T M \mathbf{z}] = \text{tr}(M\,\Sigma_s)$, we obtain:
\begin{equation}
\mathbb{E}\left[\|\delta\mathbf{x}_{r,k+1}\|^2\right] = \text{tr}(A_{rs}^T A_{rs}\,\Sigma_s) = \text{tr}(A_{rs}\,\Sigma_s\,A_{rs}^T),
\end{equation}
where the last step uses the cyclic property of the trace. Normalizing by the noise energy gives the LOCA metric stated in the main text:
\begin{equation}
F^{op}_{s \to r} = \frac{\text{tr}(A_{rs}\,\Sigma_s\,A_{rs}^T)}{\text{tr}(\Sigma_{\epsilon_r})}.
\end{equation}

\section{Two-dimensional system: full Granger derivation}
\label{si:2d_derivation}
This section derives the scalar identity $F^{GC} = F^{op}(1-\rho^2)$, Eq.~\eqref{eq:CG-op} of the main text, in three steps: the closed-form Granger causality measure, the biased restricted coefficient, and the restricted error variance.

\subsection{Closed-form Granger %
causality measure for two-state system%
}
\label{si:barnett}
For the stable and stationary two-dimensional system in the main text (Eq.~\eqref{eq:2d_system}), a closed-form expression for single-lag Granger causality %
is described in Barnett and Seth \cite{barnett2014mvgc} for the case where $\sigma_{\epsilon,x}^2 = \sigma_{\epsilon,y}^2 = 1$. In the case of arbitrary Gaussian process noise variance, the causality measure given in Eq.~\eqref{eq:GCF} has the explicit form%
\begin{equation}
F_{y\to x}^{GC} = \frac{\gamma_0^2-\gamma_1^2-\sigma_{\epsilon,x}^2\gamma_0}{\sigma_{\epsilon,x}^2\gamma_0},
\end{equation}
where
\begin{align}
\gamma_0 &= \frac{1+(1+ab)\delta}{1-a^2}\sigma_{\epsilon,x}^2,\\ \gamma_1 &= \frac{a+(a+b)\delta}{1-a^2}\sigma_{\epsilon,x}^2 , \\ \delta &= \frac{c^2}{(1-ab)(1-b^2)}\frac{\sigma_{\epsilon,y}^2}{\sigma_{\epsilon,x}^2}.
\end{align}
This expression depends on $a$ (the self-coupling of $x$), whereas the operator-theoretic metric $F^{op}_{y\to x} = c^2\,\text{var}(y)/\sigma_{\epsilon,x}^2$ does not. We have verified that this closed-form expression agrees with the general result derived in the main text (Eq.~\eqref{eq:GC_general}) for all parameter values tested.

\subsection{Least-squares derivation of the biased restricted coefficient}
\label{si:least_squares}
For the unrestricted model, we set up the least-squares problem using data sequences $\{x_k\}_{k=1}^m$ and $\{y_k\}_{k=1}^m$. Defining
\begin{align}
\bm{x}_{(1)} &= \begin{bmatrix} x_1 & x_2 & \cdots & x_{m-1} \end{bmatrix}, \\ \bm{x}_{(2)} &= \begin{bmatrix} x_2 & x_3 & \cdots & x_{m} \end{bmatrix}, \\ \bm{y}_{(1)} &= \begin{bmatrix} y_1 & y_2 & \cdots & y_{m-1} \end{bmatrix},\\
\bm{\epsilon}_{x,(1)} &= \begin{bmatrix} \epsilon_{x,1} & \epsilon_{x,2} & \cdots & \epsilon_{x,m-1} \end{bmatrix},
\end{align}
the unrestricted model $x_{k+1} = a x_k + c y_k + \epsilon_{x,k}$ can be written as
\begin{equation}
\bm{x}_{(2)} = \begin{bmatrix} a & c \end{bmatrix} \begin{bmatrix} \bm{x}_{(1)} \\ \bm{y}_{(1)} \end{bmatrix} + \bm{\epsilon}_{x,(1)},
\end{equation}
with a least-squares solution that minimizes the residual $\|\bm{\epsilon}_{x,(1)}\|$ given by
\begin{equation}
\begin{bmatrix} a & c \end{bmatrix} = \bm{x}_{(2)} \begin{bmatrix} \bm{x}_{(1)} \\ \bm{y}_{(1)} \end{bmatrix}^+,
\end{equation}
where $\cdot^+$ denotes the Moore-Penrose pseudoinverse. In the limit of large data, this converges to the true values of $a$ and $c$. (Note: if measurements contain sensor noise rather than just process noise, the least-squares fit can yield biased estimates \cite{dawson2016dmdnoise,hemati2017tls}.) For the restricted model, the estimate for $a$ (denoted $\tilde{a}$) is:
\begin{align}
\tilde{a} &= \bm{x}_{(2)} \bm{x}_{(1)}^+ = \bm{x}_{(2)} \bm{x}_{(1)}^T\left(\bm{x}_{(1)}\bm{x}_{(1)}^T\right)^{-1}.
\end{align}
In the large-data limit, this converges to
\begin{align}
\tilde{a} &= \frac{\text{cov}(x_k,x_{k+1})}{\text{var}(x)} = \frac{\text{cov}(x_k,\, ax_{k}+cy_k)}{\text{var}(x)} = a + c \frac{\text{cov}(x,y)}{\text{var}(x)},
\end{align}
as stated in the main text.

\subsection{Restricted model error variance}
\label{si:restricted_variance}
The restricted model omits $y$ and predicts $x$ from its own past alone, $x_{k+1} = \tilde{a}x_k + \tilde{\epsilon}_{x,k}$, with $\tilde{a}$ the biased coefficient derived above. Its error term is therefore the difference between the true dynamics and the restricted prediction, $\tilde{\epsilon}_{x,k} = x_{k+1} - \tilde{a}x_k = (ax_k + cy_k + \epsilon_{x,k}) - \tilde{a}x_k$. Substituting $\tilde{a} = a + c\,\text{cov}(x,y)/\text{var}(x)$ gives:
\begin{equation}
\tilde{\epsilon}_{x,k} = \epsilon_{x,k} - c\frac{\text{cov}(x,y)}{\text{var}(x)}x_k + c y_k.
\end{equation}
Its variance is:
\begin{align}
\tilde{\sigma}_{\epsilon,x}^2 &= \mathbb{E}\left[\left(\epsilon_{x,k} - c \frac{\text{cov}(x,y)}{\text{var}(x)}x_k + c y_k\right)^2\right]\\ &= \mathbb{E}\left[\epsilon_{x,k}^2 + c^2 \frac{\text{cov}(x,y)^2}{\text{var}(x)^2}x_k^2 + c^2 y_k^2 - 2c^2\frac{\text{cov}(x,y)}{\text{var}(x)}x_k y_k\right] \\ &= \sigma_{\epsilon,x}^2 + c^2\text{var}(y) + c^2 \frac{\text{cov}(x,y)^2}{\text{var}(x)} - 2c^2\frac{\text{cov}(x,y)^2}{\text{var}(x)} \\ &= \sigma_{\epsilon,x}^2 + c^2\text{var}(y) - c^2 \frac{\text{cov}(x,y)^2}{\text{var}(x)},
\end{align}
where we have used the fact that $\epsilon_{x,k}$ is uncorrelated with both $x_k$ and $y_k$. Substituting into the Granger causality measure %
 definition gives Eq.~\eqref{eq:GC_general} of the main text. The required covariances can also be computed without data by solving the discrete Lyapunov equation $\Sigma = A\Sigma A^T + Q$ \cite{lutkepohl2005new}.

\section{N-dimensional generalization: full derivation}
\label{si:ndim}
This section generalizes the scalar identity to $N$ dimensions and establishes the conditioning ordering of Eq.~\eqref{eq:ordering} of the main text. The two central objects are the restricted coefficient matrix and the conditional source covariance.

We derive the three-partition results stated in the main text. Let $s$, $r$, and $o$ denote disjoint index sets for source, receiver (target), and other (conditioning) variables, with $s\cup r\cup o=\{1,\ldots,N\}$. Define the combined retained predictor $\mathbf{z}_k = [\mathbf{x}_{r,k};\,\mathbf{x}_{o,k}]$, with $\Sigma_z = \mathbb{E}[\mathbf{z}_k\mathbf{z}_k^T]$ and $\Sigma_{sz} = \mathbb{E}[\mathbf{x}_{s,k}\mathbf{z}_k^T]$. The two-partition result (no conditioning variables) follows by setting $\mathbf{x}_o=\emptyset$, so $\mathbf{z}_k=\mathbf{x}_{r,k}$, $\Sigma_z=\Sigma_r$, and $\Sigma_{sz}=\Sigma_{sr}$.

\subsection{Restricted model coefficient matrix}
\label{si:ares}
The unrestricted receiver equation is $\mathbf{x}_{r,k+1} = A_{rz}\mathbf{z}_k + A_{rs}\mathbf{x}_{s,k} + \boldsymbol{\epsilon}_{r,k}$. The restricted model predicts $\mathbf{x}_{r,k+1}$ using only the retained variables, $\mathbf{x}_{r,k+1} = \tilde{A}_{rz}\mathbf{z}_k + \tilde{\boldsymbol{\epsilon}}_{r,k}$. The coefficient minimizes the mean squared prediction error $\mathbb{E}\|\mathbf{x}_{r,k+1} - \tilde{A}_{rz}\mathbf{z}_k\|^2$, whose normal equations require the residual to be uncorrelated with the regressors, $\mathbb{E}[(\mathbf{x}_{r,k+1} - \tilde{A}_{rz}\mathbf{z}_k)\mathbf{z}_k^T] = \mathbf{0}$. This is the population (large-data) analog of the sample least-squares fit of Section~\ref{si:2d_derivation}. Solving for the coefficient gives:
\begin{equation}
\tilde{A}_{rz} = \mathbb{E}[\mathbf{x}_{r,k+1}\mathbf{z}_k^T]\,\Sigma_z^{-1}.
\end{equation}
Substituting the unrestricted model for $\mathbf{x}_{r,k+1}$ and using $\mathbb{E}[\boldsymbol{\epsilon}_{r,k}\mathbf{z}_k^T]=\mathbf{0}$:
\begin{align}
\mathbb{E}[\mathbf{x}_{r,k+1}\mathbf{z}_k^T] &= \mathbb{E}\left[\left(A_{rz}\mathbf{z}_k + A_{rs}\mathbf{x}_{s,k} + \boldsymbol{\epsilon}_{r,k}\right)\mathbf{z}_k^T\right] \nonumber\\
&= A_{rz}\,\mathbb{E}[\mathbf{z}_k\mathbf{z}_k^T] + A_{rs}\,\mathbb{E}[\mathbf{x}_{s,k}\mathbf{z}_k^T] + \mathbb{E}[\boldsymbol{\epsilon}_{r,k}\mathbf{z}_k^T] \nonumber\\
&= A_{rz}\Sigma_z + A_{rs}\Sigma_{sz}.
\end{align}
Therefore:
\begin{equation}
\tilde{A}_{rz} = A_{rz} + A_{rs}\Sigma_{sz}\Sigma_z^{-1}.
\end{equation}
This is Eq.~\eqref{eq:Atilde_rz} of the main text. In words, the restricted fit does not simply remove the source from the model. It absorbs the component of the source's influence that is linearly predictable from the retained variables. The bias term $A_{rs}\Sigma_{sz}\Sigma_z^{-1}$ generalizes $c\,\text{cov}(x,y)/\text{var}(x)$ from the scalar case, now projecting the source onto the full retained predictor set $\mathbf{z}_k$.

\subsection{Restricted error covariance and conditional source covariance}
\label{si:restricted_covariance}
The restricted residual is $\tilde{\boldsymbol{\epsilon}}_{r,k} = \mathbf{x}_{r,k+1} - \tilde{A}_{rz}\mathbf{z}_k$. Substituting the unrestricted model and $\tilde{A}_{rz}$:
\begin{align}
\tilde{\boldsymbol{\epsilon}}_{r,k} &= A_{rz}\mathbf{z}_k + A_{rs}\mathbf{x}_{s,k} + \boldsymbol{\epsilon}_{r,k} - \left(A_{rz} + A_{rs}\Sigma_{sz}\Sigma_z^{-1}\right)\mathbf{z}_k \nonumber\\
&= A_{rs}\!\left(\mathbf{x}_{s,k} - \Sigma_{sz}\Sigma_z^{-1}\mathbf{z}_k\right) + \boldsymbol{\epsilon}_{r,k} = A_{rs}\mathbf{x}_{s,k}^* + \boldsymbol{\epsilon}_{r,k},
\end{align}
where $\mathbf{x}_{s,k}^* \equiv \mathbf{x}_{s,k} - \Sigma_{sz}\Sigma_z^{-1}\mathbf{z}_k$ is the component of the source orthogonal to the full retained predictor $\mathbf{z}_k$. Since $\mathbf{x}_{s,k}^*$ is a function of states at time $k$, it is uncorrelated with $\boldsymbol{\epsilon}_{r,k}$. We now compute the covariance of $\mathbf{x}_{s,k}^*$. The states $(\mathbf{z}_k, \mathbf{x}_{s,k})$ have joint covariance $\begin{bmatrix}\Sigma_z & \Sigma_{zs}\\ \Sigma_{sz} & \Sigma_s\end{bmatrix}$, with blocks $\Sigma_s = \mathbb{E}[\mathbf{x}_{s,k}\mathbf{x}_{s,k}^T]$, $\Sigma_{sz} = \mathbb{E}[\mathbf{x}_{s,k}\mathbf{z}_k^T]$, and $\Sigma_{zs} = \Sigma_{sz}^T$. Expanding the covariance of $\mathbf{x}_{s,k}^*$ term by term:
\begin{align}
\Sigma_{s|(r,o)} &\equiv \mathbb{E}[\mathbf{x}_{s,k}^*(\mathbf{x}_{s,k}^*)^T] \nonumber\\
&= \mathbb{E}\!\left[\left(\mathbf{x}_{s,k} - \Sigma_{sz}\Sigma_z^{-1}\mathbf{z}_k\right)\left(\mathbf{x}_{s,k} - \Sigma_{sz}\Sigma_z^{-1}\mathbf{z}_k\right)^T\right] \nonumber\\
&= \Sigma_s - \Sigma_{sz}\Sigma_z^{-1}\Sigma_{zs} - \Sigma_{sz}\Sigma_z^{-1}\Sigma_{zs} + \Sigma_{sz}\Sigma_z^{-1}\Sigma_z\Sigma_z^{-1}\Sigma_{zs} \nonumber\\
&= \Sigma_s - \Sigma_{sz}\Sigma_z^{-1}\Sigma_{zs},
\end{align}
where the middle line uses the four second moments of $(\mathbf{x}_{s,k}, \mathbf{z}_k)$ together with the symmetry of the covariance $\Sigma_z$ (so $\Sigma_z^{-1}$ is also symmetric), and the last two terms combine because $\Sigma_z^{-1}\Sigma_z = \mathbf{I}$. The result is the Schur complement of the $\Sigma_z$ block in the joint covariance above \cite{zhang2005schur}.
Expanding $\Sigma_z$, $\Sigma_{sz}$, and $\Sigma_{zs}$ in terms of the $r$, $s$, $o$ covariance blocks (where $\Sigma_{ro}=\mathbb{E}[\mathbf{x}_{r,k}\mathbf{x}_{o,k}^T]$, $\Sigma_{so}=\mathbb{E}[\mathbf{x}_{s,k}\mathbf{x}_{o,k}^T]$, etc.):
\begin{equation}\label{si:eq:sigma_s_given_ro}
\Sigma_{s|(r,o)} = \Sigma_s - \begin{bmatrix}\Sigma_{sr} & \Sigma_{so}\end{bmatrix}\begin{bmatrix}\Sigma_{rr} & \Sigma_{ro}\\\Sigma_{or} & \Sigma_{oo}\end{bmatrix}^{-1}\begin{bmatrix}\Sigma_{rs}\\\Sigma_{os}\end{bmatrix}.
\end{equation}
Thus, $\Sigma_{s|(r,o)}$ is the covariance remaining in $\mathbf{x}_s$ after subtracting the best linear prediction from the receiver variables $\mathbf{x}_r$ and conditioning variables $\mathbf{x}_o$. Only this linearly unpredictable part of the source is attributed to the source in Granger causality. %
The restricted error covariance is therefore:
\begin{equation}\label{si:eq:Sigma_tilde}
\Sigma_{\tilde{\epsilon}_r} = A_{rs}\,\Sigma_{s|(r,o)}\,A_{rs}^T + \Sigma_{\epsilon_r},
\end{equation}
and the conditional Granger causality measure %
and $N$-dimensional correlation penalty follow directly as stated in the main text.

\subsection{Two-partition reduction}
\label{si:two_partition_reduction}
The two-partition result is recovered by taking the conditioning set to be empty, $\mathbf{x}_o=\emptyset$, so that $\mathbf{z}_k=\mathbf{x}_{r,k}$, $\Sigma_z=\Sigma_r$, $\Sigma_{sz}=\Sigma_{sr}$, and $\Sigma_{zs}=\Sigma_{rs}$. Then:
\begin{equation}
\Sigma_{s|(r,o)}\;\longrightarrow\;\Sigma_{s|r} = \Sigma_s - \Sigma_{sr}\Sigma_r^{-1}\Sigma_{rs},
\end{equation}
and the conditional Granger causality measure reduces to $F_{s\to r}^{GC} = \mathrm{tr}(A_{rs}\Sigma_{s|r}A_{rs}^T)/\mathrm{tr}(\Sigma_{\epsilon_r})$, with bias $F_{s\to r}^{op}-F_{s\to r}^{GC} = \mathrm{tr}(A_{rs}\Sigma_{sr}\Sigma_r^{-1}\Sigma_{rs}A_{rs}^T)/\mathrm{tr}(\Sigma_{\epsilon_r})\geq 0$. In the following, $M \succeq \mathbf{0}$ denotes that the symmetric matrix $M$ is positive semidefinite, $\mathbf{v}^T M \mathbf{v} \ge 0$ for every vector $\mathbf{v}$. Correspondingly, $A \preceq B$ means $B - A \succeq \mathbf{0}$. The monotonic ordering $\Sigma_{s|(r,o)}\preceq\Sigma_{s|r}\preceq\Sigma_s$ follows from the positive semidefiniteness of each successive Schur complement correction:
\begin{align}
\Sigma_s - \Sigma_{s|r} &= \Sigma_{sr}\Sigma_r^{-1}\Sigma_{rs} \succeq 0,\\
\Sigma_{s|r} - \Sigma_{s|(r,o)} &= \Sigma_{sz}\Sigma_z^{-1}\Sigma_{zs} - \Sigma_{sr}\Sigma_r^{-1}\Sigma_{rs} \succeq 0,
\end{align}
where the second inequality follows from the fact that conditioning on a strictly larger set ($\mathbf{z}$ vs.\ $\mathbf{x}_r$) can only reduce the residual variance. In words, adding conditioning variables can only reduce the source covariance left unexplained by the retained predictors. With the same full-state propagator block $A_{rs}$, this can only decrease Granger causality relative to the operator response. This gives the monotonic ordering of the causality measures %
stated in the main text:
\begin{equation}
F_{s\to r|o}^{GC} \leq F_{s\to r}^{GC} \leq F_{s\to r}^{op}.
\end{equation}

\emph{Remark on two-partition vs.\ classical pairwise Granger.} The statistic $F^{GC}_{s\to r}$ defined here (the $\mathbf{x}_o=\emptyset$ reduction of Eq.~\eqref{si:eq:Sigma_tilde}) uses the full $N$-variable propagator block $A_{rs}$, not a coefficient estimated from a reduced VAR fitted only to $(\mathbf{x}_r, \mathbf{x}_s)$. The two-partition label refers solely to the covariance weighting $\Sigma_{s|r}$. The regression model remains the full $N$-variable unrestricted VAR throughout. This distinguishes $F^{GC}_{s\to r}$ from the classical pairwise Granger statistic, which uses a biased $\hat{A}_{rs}$ obtained from a reduced VAR that omits $\mathbf{x}_o$ entirely (see Section~\ref{si:full_state}).

\subsection{Scalar bivariate reduction}
\label{si:scalar_reduction}
For a single receiver $x_i$, single source $x_j$, and no conditioning variables ($\mathbf{x}_o=\emptyset$), the matrix quantities reduce to scalars: $A_{rs}\to a_{ij}$, $\Sigma_s\to\mathrm{var}(x_j)$, $\Sigma_z\to\mathrm{var}(x_i)$, $\Sigma_{sz}\to\mathrm{cov}(x_j,x_i)$, $\Sigma_{\epsilon_r}\to\sigma_{\epsilon,i}^2$. The partial covariance reduces to $\Sigma_{s|(r,o)}\to\mathrm{var}(x_j)-\mathrm{cov}(x_j,x_i)^2/\mathrm{var}(x_i)$, and the Granger causality measure %
recovers the two-dimensional result of the main text:
\begin{equation}
F_{j\to i}^{GC} = \frac{a_{ij}^2}{\sigma_{\epsilon,i}^2}\!\left(\mathrm{var}(x_j) - \frac{\mathrm{cov}(x_i,x_j)^2}{\mathrm{var}(x_i)}\right) = F^{op}_{j\to i}\!\left(1-\rho_{ij}^2\right).
\end{equation}
For the scalar three-partition case (single receiver $x_i$, single source $x_j$, conditioning variables $\mathbf{x}_o$ non-empty), the conditional source variance is the scalar $\Sigma_{s|(r,o)} = \mathrm{var}(x_j) - \Sigma_{sz}\Sigma_z^{-1}\Sigma_{zs}$, where $\mathbf{z}=(x_i,\mathbf{x}_o^T)^T$, and the conditional Granger causality measure is $F_{j\to i|o}^{GC} = a_{ij}^2\Sigma_{s|(r,o)}/\sigma_{\epsilon,i}^2$. The penalty $\mathcal{P}_{j\to i|o} = 1 - \Sigma_{s|(r,o)}/\mathrm{var}(x_j)$ is in general the multiple $R^2$ of $x_j$ regressed jointly on $\mathbf{z}$, not simply the pairwise $\rho^2(x_i,x_j)$. The two coincide only in the bivariate case, where $\mathbf{z}$ reduces to $x_i$ alone.

\subsection{Remark: scope of the fully observed state assumption}
\label{si:full_state}
The monotonic ordering $F^{GC}_{s\to r|o}\leq F^{GC}_{s\to r}\leq F^{op}_{s\to r}$ holds because all three statistics share a single propagator block $A_{rs}$, obtained by fitting one full $N$-variable VAR to all variables simultaneously. Only the source covariance matrix changes across the three comparisons ($\Sigma_s\to\Sigma_{s|r}\to\Sigma_{s|(r,o)}$). Since each successive Schur complement correction is positive semidefinite, the covariances are ordered and so are the $F$ values.%

This structure breaks down when the full modeled state vector is not observed. If a subset of variables $\mathbf{x}_o$ is unobserved and separate reduced VAR models are fit for each partitioning, the propagator block $A_{rs}$ itself changes between comparisons, absorbing the influence of the omitted variables. Two competing effects then arise: (i)~\emph{coefficient bias}, in which $\hat{A}_{rs}$ deviates from the full-model $A_{rs}$ in a direction that depends on the signs of the omitted pathways, and (ii)~\emph{residual inflation}, in which the restricted model error variance $\mathrm{tr}(\hat{\Sigma}_{\epsilon_r})$ is always larger than $\mathrm{tr}(\Sigma_{\epsilon_r})$. These competing effects mean the monotonic ordering cannot be guaranteed, and the Granger statistic may lie above or below the operator response in ways that are not predictable from the correlation structure alone. The two applications in the main text, an 80-variable observed VAR for fMRI and an 8-cell Burgers advection chain, both operate with all model variables observed, so the ordering holds exactly by construction, as verified numerically with zero violations across all directed pairs.

\subsection{Relation to full-history Granger causality}
\label{si:fullhistory}
The Granger statistics used in this work are single-lag quantities: both the unrestricted and restricted models predict $\mathbf{x}_{r,k+1}$ from the state at time $k$ only. Standard (Geweke) Granger causality instead conditions each model on the entire past of its variable set \cite{geweke1982measurement}, and the reduced process obtained by omitting the source is generically of infinite VAR order \cite{barnett2015granger}, so the two statistics differ in general. The upper-bound structure nevertheless extends.

\textbf{Proposition.} Let the fully observed modeled state follow the VAR(1) model of Eq.~\eqref{eq:discrete_system} of the main text with all variables observed, and let $F^{GC,\infty}_{s\to r|o}$ denote the full-history analog of the trace-based variance-increment statistic, in which the unrestricted model predicts $\mathbf{x}_{r,k+1}$ from the full past of all variables and the restricted model predicts it from the full past of $(\mathbf{x}_r,\mathbf{x}_o)$. Then
\begin{equation}
F^{GC,\infty}_{s\to r|o} \;\leq\; F^{GC}_{s\to r|o} \;=\; F^{op}_{s\to r}\left(1-\mathcal{P}_{s\to r|o}\right) \;\leq\; F^{op}_{s\to r}.
\end{equation}

Physically, $F^{GC,\infty}_{s\to r|o}$ is the statistic computed in conventional practice, where the restricted model may use unlimited lags. The omitted source leaves delayed, filtered copies of its activity in the history of the retained variables, so a longer memory lets the restricted model reconstruct more of the source's influence without observing the source. The Granger statistic can therefore only decrease, and the penalty only grow, relative to the one-lag computation.

\textit{Proof.} The proof combines two standard facts \cite{lutkepohl2005new}: the Markov property of the VAR(1) process and the monotonic decrease of least-squares prediction-error covariance as the conditioning set grows. Because the modeled VAR(1) state is Markov, the optimal linear predictor of $\mathbf{x}_{r,k+1}$ from the full past of all variables coincides with the one-lag predictor, so the unrestricted residual covariance $\Sigma_{\epsilon_r}$ is unchanged by adding history. The optimal linear predictor from the full past of $(\mathbf{x}_r,\mathbf{x}_o)$ projects onto a subspace that contains the one-lag restricted regressors, so its residual covariance satisfies $\Sigma^{\infty}_{\tilde{\epsilon}_r} \preceq \Sigma_{\tilde{\epsilon}_r}$ in the positive semidefinite order. The trace-based statistic of Eq.~\eqref{eq:Granger-N-dim-F} of the main text is increasing in $\mathrm{tr}\,\Sigma_{\tilde{\epsilon}_r}$ at fixed $\Sigma_{\epsilon_r}$, and the log-determinant (Geweke/transfer entropy) statistic of Section~\ref{si:te} is increasing in $\Sigma_{\tilde{\epsilon}_r}$ in the positive semidefinite order. Since full-history conditioning shrinks the restricted covariance while leaving $\Sigma_{\epsilon_r}$ unchanged, both inequalities follow. \hfill$\square$

Consequently, the single-lag trace penalty $\mathcal{P}_{s\to r|o}$ is a lower bound on the attenuation of the corresponding full-history trace analog. Applying the same projection argument to the log-determinant statistic yields the corresponding result for the conventional Geweke causality test. Hence the one-lag log-determinant penalty $\mathcal{P}_{TE}$ is a lower bound on its full-history attenuation. An exact expression for the full-history gap would require the steady-state filtering covariance of the reduced (VARMA) process and is left for future work.

\section{Connection to transfer entropy}
\label{si:te}
\label{si:transfer_entropy}
This section carries the correlation penalty over to transfer entropy. For jointly Gaussian processes, transfer entropy and Granger causality are equivalent, so the penalty applies on the information-theoretic scale as well. We first review the Gaussian equivalence between Granger causality and transfer entropy, then derive the multivariate generalization of the operator-theoretic transfer entropy bound.

\subsection{Gaussian equivalence of Granger causality and transfer entropy}
\label{si:te_gaussian}
The one-lag transfer entropy considered in this work, from $y$ to $x$, is the conditional mutual information \cite{schreiber2000te}
\begin{equation}\label{eq:TE_def}
TE_{y \to x} = I(x_{k+1}\,;\, y_k \mid x_k) = \mathbb{E}\!\left[\log_2\frac{p(x_{k+1}\mid x_k, y_k)}{p(x_{k+1}\mid x_k)}\right],
\end{equation}
where the expectation is taken with respect to the joint distribution $p(x_{k+1}, x_k, y_k)$. For jointly Gaussian processes, all conditional distributions are Gaussian, and the conditional differential entropy reduces to a function of the conditional variance only. For any conditioning set $\mathcal{C}$ (here either $\{x_k\}$ or $\{x_k, y_k\}$):
\begin{equation}
h(x_{k+1}\mid \mathcal{C}) = \tfrac{1}{2}\log_2\!\left[2\pi e\,\text{var}(x_{k+1}\mid \mathcal{C})\right].
\end{equation}
The variances $\text{var}(x_{k+1}\mid x_k)$ and $\text{var}(x_{k+1}\mid x_k, y_k)$ are precisely the residual variances of the restricted and unrestricted regression models defined in the main text, $\tilde{\sigma}_{\epsilon,x}^2$ and $\sigma_{\epsilon,x}^2$ respectively. Substituting:
\begin{equation}\label{eq:TE_var_ratio}
TE_{y \to x} = \tfrac{1}{2}\log_2\frac{\tilde{\sigma}_{\epsilon,x}^2}{\sigma_{\epsilon,x}^2}.
\end{equation}
The Granger causality measure %
 from Eq.~\eqref{eq:GCF} of the main text is the relative increase in residual variance, $F^{GC}_{y\to x} = \left[\tilde{\sigma}_{\epsilon,x}^2 - \sigma_{\epsilon,x}^2\right]/\sigma_{\epsilon,x}^2$, so the variance ratio in Eq.~\ref{eq:TE_var_ratio} is simply $1 + F^{GC}_{y\to x}$, giving the equivalence stated in the main text:
\begin{equation}\label{eq:TE_GC_si}
TE_{y\to x} = \tfrac{1}{2}\log_2\!\left(1 + F^{GC}_{y\to x}\right).
\end{equation}
This relation is due to Barnett, Barrett \& Seth \cite{barnett2009granger}. We use $\log_2$ throughout so that TE is measured in bits.

\subsection{Multivariate operator-theoretic transfer entropy}
\label{si:te_multivariate}
For multivariate Gaussian sources $\mathbf{x}_s$ and receivers $\mathbf{x}_r$, transfer entropy generalizes from a variance ratio to a determinant ratio of residual covariance matrices:
\begin{equation}\label{eq:TE_multivariate}
TE_{s \to r|o} = \tfrac{1}{2}\log_2\frac{\left|\Sigma_{\tilde{\epsilon}_r}\right|}{\left|\Sigma_{\epsilon_r}\right|}.
\end{equation}
Substituting the restricted error covariance derived in Section~\ref{si:restricted_covariance}, $\Sigma_{\tilde{\epsilon}_r} = A_{rs}\,\Sigma_{s|(r,o)}\,A_{rs}^T + \Sigma_{\epsilon_r}$, gives:
\begin{equation}\label{eq:TE_GC_multivariate}
TE_{s \to r|o} = \tfrac{1}{2}\log_2\left|\,\mathbf{I} + \Sigma_{\epsilon_r}^{-1}\,A_{rs}\,\Sigma_{s|(r,o)}\,A_{rs}^T\right|.
\end{equation}
The corresponding operator-theoretic upper bound, obtained by replacing the conditional covariance $\Sigma_{s|(r,o)}$ with the full source covariance $\Sigma_s$, is:
\begin{equation}\label{eq:TE_op_multivariate}
TE^{op}_{s \to r} \equiv \tfrac{1}{2}\log_2\left|\,\mathbf{I} + \Sigma_{\epsilon_r}^{-1}\,A_{rs}\,\Sigma_s\,A_{rs}^T\right|.
\end{equation}
To establish the inequality $TE_{s\to r|o} \leq TE^{op}_{s\to r}$, first note that, for either $B=A_{rs}\Sigma_{s|(r,o)}A_{rs}^T$ or $B=A_{rs}\Sigma_sA_{rs}^T$,
\begin{equation}
\det\!\left(\mathbf{I}+\Sigma_{\epsilon_r}^{-1}B\right)
=
\det\!\left(
\mathbf{I}+\Sigma_{\epsilon_r}^{-1/2}
B
\Sigma_{\epsilon_r}^{-1/2}
\right).
\end{equation}
Thus, the two log-determinants may equivalently be evaluated in
noise-whitened coordinates. Since $\Sigma_s-\Sigma_{s|(r,o)}
=\Sigma_{sz}\Sigma_z^{-1}\Sigma_{zs}\succeq\mathbf{0}$, we have $\Sigma_{s|(r,o)}\preceq\Sigma_s$, and the corresponding noise-whitened response covariances have the same positive-semidefinite ordering. The matrices inside the determinants are therefore symmetric positive definite, and monotonicity of $\log\det$ gives $TE_{s\to r|o}\leq TE^{op}_{s\to r}$. 
Equality holds when $A_{rs}\,\Sigma_{sz}\Sigma_z^{-1}\Sigma_{zs}\,A_{rs}^T = \mathbf{0}$. Thus $\Sigma_{sz} = \mathbf{0}$ is sufficient, but in multiple dimensions equality may also occur when the correlation-explained source subspace does not couple to the receiver through $A_{rs}$. 
The multivariate correlation penalty quantifying the gap is:
\begin{equation}\label{eq:penalty_TE}
\mathcal{P}^{TE}_{s \to r|o} = 1 - \frac{TE_{s\to r|o}}{TE^{op}_{s\to r}} = 1 - \frac{\log_2\!\left|\mathbf{I} + \Sigma_{\epsilon_r}^{-1}A_{rs}\Sigma_{s|(r,o)}A_{rs}^T\right|}{\log_2\!\left|\mathbf{I} + \Sigma_{\epsilon_r}^{-1}A_{rs}\Sigma_{s}A_{rs}^T\right|}.
\end{equation}
In the scalar limit ($\mathbf{x}_o=\emptyset$, single source and receiver), the determinants reduce to scalars, $|\mathbf{I} + \cdot| \to 1 + F^{op}(1-\rho^2)$ and $1 + F^{op}$ respectively, recovering the bound in Eq.~\eqref{eq:TE_GC_relation} of the main text.

\textit{Numerical TE estimation for Fig.~\ref{fig:2Dsys}(b).} The numerical markers use a plug-in conditional mutual information estimate with quantile bins ($128\times 128\times 32$ for $x_{k+1}\times x_k\times y_k$) on $10^7$ samples after a $2\times 10^4$ burn-in. A shuffle correction subtracts the mean of 20 source-shuffled surrogate estimates from the raw estimate, removing finite-bin bias \cite{marschinski2002transfer}. Interventional ($TE^{op}$) markers convert the raw observational estimate through the Gaussian variance-ratio factor of Eq.~\eqref{eq:TE_bound} of the main text. Binned estimators converge slowly for strongly autocorrelated signals, producing the visible marker deviations near $a \approx \pm 1$ while the analytical curves remain exact.

\section{Linearized Burgers advection chain: derivation and validation}
\label{si:burgers}
This section derives the finite-horizon statistics for the advection chain of Fig.~\ref{fig:burgers} of the main text and validates the linearized chain against nonlinear Burgers solutions.

\subsection{PDE linearization and discrete chain}
\label{si:burgers_derivation}
The viscous Burgers equation with mean flow $U_0$ is
\begin{equation}
\frac{\partial u}{\partial t} + u\frac{\partial u}{\partial x} = \nu\frac{\partial^2 u}{\partial x^2}.
\end{equation}
Writing $u = U_0 + q$ and neglecting the quadratic term $q\partial_x q$ gives the linear advection--diffusion equation $\partial_t q + U_0 \partial_x q = \nu \partial_{xx} q$. In the convection-dominated limit relevant here, first-order upwind discretization on a uniform grid with spacing $\Delta x$ and time step $\Delta t$, with the stochastic forcing $\eta$ carried through as the cell noise $\epsilon_{m,k}$, yields the open downstream chain
\begin{equation}
x_{m,k+1} = a\,x_{m,k} + c\,x_{m-1,k} + \epsilon_{m,k}, \qquad m=1,\ldots,N,
\end{equation}
with $a = 1 - U_0\Delta t/\Delta x$ and $c = U_0\Delta t/\Delta x$ (Courant number). For the parameter values in the main text ($U_0=2$, $\Delta x=1$, $\Delta t=0.1$) this gives $a=0.8$, $c=0.2$, CFL $= 0.2 < 1$ (scheme stable). In matrix form, $\mathbf{x}_{k+1} = A\mathbf{x}_k + \boldsymbol{\epsilon}_k$ with $A = aI + cS$, where $S$ is the $N\times N$ lower-shift matrix ($[S]_{m,m-1}=1$, all other entries zero).

\subsection{Finite-horizon LOCA and the travel-time peak}
\label{si:burgers_horizon}
\label{si:burgers_loca}
Since $I$ and $S$ commute, the $h$-step propagator factors as a binomial series (this expansion underlies the travel-time peak quoted in the main text):
\begin{equation}
A^h = (aI + cS)^h = \sum_{\ell=0}^{\min(h,N-1)}\binom{h}{\ell}a^{h-\ell}c^\ell S^\ell.
\end{equation}
The $(\ell)$-th term shifts a signal downstream by $\ell$ cells. For source cell 1 and receiver cell 4 ($d=3$ steps downstream), only the $\ell=3$ term contributes:
\begin{equation}
[A^h]_{4,1} = \binom{h}{3}a^{h-3}c^3.
\end{equation}
Physically, the response grows while the disturbance is still arriving at the receiver and decays once it has passed, so the sequence rises and then falls. The discrete sequence $g_h = \binom{h}{3}a^{h-3}c^3$ satisfies $g_{h+1}/g_h = a\,(h+1)/(h-2)$. For $a=0.8$ and $d=3$ this ratio equals unity at $h=14$, so the response has equal maxima at $h=14$ and $h=15$ and decreases thereafter. We use $h=15$, which coincides with the continuum travel time $d\Delta x/U_0 = 1.5$\,s. In the limit $\Delta t\to 0$, $a \to 1 - U_0\Delta t/\Delta x$ and the discrete peak converges to the continuum travel time. The choice between the two equal maxima does not affect the penalty, because the common factor $[A^h]_{41}$ cancels from the penalty ratio.

The stationary state covariance $\Sigma$ is obtained from the discrete Lyapunov equation $\Sigma = A\Sigma A^T + Q$, with noise covariance $Q = \mathrm{diag}(10,1,\ldots,1)$ (upstream-dominated). The finite-horizon statistics follow from Eqs.~\eqref{eq:Fop_group} and~\eqref{eq:Granger-N-dim-F} of the main text with $A$ replaced by $A^h$. These are finite-horizon extensions of the operator-weighted one-lag statistics, retaining the fixed one-step noise normalization $\Sigma_{\epsilon,44}$ (the $(4,4)$ entry of $\Sigma_\epsilon$, the one-step noise variance of receiver cell 4). They are not conventional $h$-step prediction-error Granger statistics, whose unrestricted error would contain the accumulated innovation covariance $\Omega_h = \sum_{\ell=0}^{h-1} A^\ell Q (A^\ell)^T$. The penalty ratio is unaffected by this common normalization:
\begin{align}
F^{op}_{s\to r}(h) &= \frac{[A^h]_{4,1}^2\,\Sigma_{11}}{\Sigma_{\epsilon,44}}, \\
F^{GC}_{s\to r|o}(h) &= \frac{[A^h]_{4,1}^2\,\Sigma_{s|(r,o)}}{\Sigma_{\epsilon,44}},
\end{align}
The two statistics share the same propagator factor and noise floor. They differ only in the source covariance, unconditional for $F^{op}$ and conditional for $F^{GC}$. Here $\Sigma_{s|(r,o)}$ is the Schur-complement residual source variance from Eq.~\eqref{si:eq:sigma_s_given_ro}, with $s=\{1\}$, $r=\{4\}$, and $o=\{2,3,5,6,7,8\}$ (all cells except source and receiver). The noise variance $\Sigma_{\epsilon,44} = Q_{44} = 1$ for $m>1$. Because the upstream noise $Q_{11}=10$ drives correlated copies of $x_1$'s signal into all downstream cells, conditioning on $o$ removes most predictive information about cell 1, yielding $\Sigma_{s|(r,o)} \ll \Sigma_{11}$ and a substantial penalty. At the peak $h=15$, the ratio $\Sigma_{s|(r,o)}/\Sigma_{11}$ equals $1 - \mathcal{P}_{s\to r|o} = 0.634$, giving the $36.6\%$ penalty.

\subsection{$N$-sensitivity}
\label{si:burgers_nsensitivity}
To assess robustness, we repeat the analysis for chain lengths $N\in\{4,5,6,8,10,12,16,24\}$ with all other parameters fixed. The penalty $\mathcal{P}_{3p}$ saturates at $N=6$ and remains constant to four significant figures for all larger chains (Table~\ref{tab:burgers_nsens}). The selected peak horizon $h^*=15$ (one of the two equal discrete maxima; Section~\ref{si:burgers_horizon}) is likewise independent of $N$. Both results confirm that the $36.6\%$ penalty is determined by the local correlation between source cell 1 and the nearest downstream conditioning cells (cells 2 and 3), with cells further downstream contributing negligible additional explanatory power.

\begin{table}[h]
\centering
\caption{Three-partition penalty $\mathcal{P}_{3p}$ at the selected peak horizon $h^*=15$, one of the two equal discrete maxima ($t^*=1.5$\,s $= d\Delta x/U_0$), for the Burgers chain as a function of chain length $N$. Source: cell 1; receiver: cell 4; conditioning set $o$: all remaining cells. Parameters: $a=0.8$, $c=0.2$, $Q=\mathrm{diag}(10,1,\ldots,1)$. Both $h^*$ and $\mathcal{P}_{3p}$ are independent of $N$ for $N\geq 6$, confirming that the penalty is set by the local source--conditioning correlation structure and not by chain length.}
\label{tab:burgers_nsens}
\begin{tabular}{cccc}
\toprule
$N$ & $h^*$ & $F^{GC}_{3p}/F^{op}$ & $\mathcal{P}_{3p}$ \\
\midrule
4  & 15 & 0.6346 & 0.3654 \\
5  & 15 & 0.6344 & 0.3656 \\
6  & 15 & 0.6343 & 0.3657 \\
8  & 15 & 0.6343 & 0.3657 \\
10 & 15 & 0.6343 & 0.3657 \\
12 & 15 & 0.6343 & 0.3657 \\
16 & 15 & 0.6343 & 0.3657 \\
24 & 15 & 0.6343 & 0.3657 \\
\bottomrule
\end{tabular}
\end{table}

\subsection{Nonlinear Burgers validation}
\label{si:burgers_nonlinear}
To assess the small-amplitude assumption, we compare the linearized chain against numerical solutions of the full nonlinear Burgers PDE for two amplitude ratios: $A_0/U_0 = 0.02$ (small amplitude) and $A_0/U_0 = 0.2$ (finite amplitude), with all other parameters ($U_0$, $\Delta x$, $\Delta t$, source location, receiver location) identical to the LOCA analysis. The perturbation is a Gaussian pulse $q_0(x) = A_0\exp[-(x-x_0)^2/w^2]$ with width $w = 0.8$. The inviscid shock-formation time is $t_{sh} = -1/\min_x q_0'(x) = w\,e^{1/2}/(\sqrt{2}\,A_0)$, where the uniform base $U_0$ adds only a constant to $u_0$ and does not affect $t_{sh}$. This gives $t_{sh} = 23.3$\,s for the small-amplitude case and $t_{sh} = 2.33$\,s for the finite-amplitude case. The validation window is $0 \leq t \leq 1.98$\,s, chosen to bracket the linear travel time $t^* = 1.5$\,s while remaining within one shock time of the finite-amplitude simulation.

For $A_0/U_0 = 0.02$, the shock forms far outside the validation window ($t_{sh} = 23.3$\,s $\gg 1.98$\,s) and the nonlinear solution is indistinguishable from the linearized prediction throughout: the disturbance arrives at the receiver at $t^* = 1.5$\,s as predicted. For $A_0/U_0 = 0.2$, nonlinear self-advection steepens the wavefront, causing the disturbance to arrive at the receiver slightly earlier than the linear travel time. The shock forms at $t_{sh} = 2.33$\,s, just after the validation window closes. The linearized propagator $A$ therefore slightly underestimates the propagated response at horizons near $t^*$ for finite-amplitude disturbances, and overestimates it at longer horizons once the nonlinear shock has dissipated. For the parameter regime of the LOCA analysis ($A_0/U_0 = 0.02$), the linearization is accurate throughout the relevant horizon range. A complementary and fully nonlinear treatment, using exact Cole--Hopf realizations of finite-amplitude acoustic Burgers propagation that span the range from linear acoustics to mature $N$-waves, is given in Section~\ref{si:acoustic}.

\section{HCP resting-state fMRI: data and methods}
\label{si:hcp_methods}
This section documents the dataset, model fitting, statistics, and diagnostics behind the brain-network results of Fig.~\ref{fig:hcp} of the main text.

\subsection{Dataset}
We analyze resting-state BOLD time series from the Human Connectome Project (HCP) 1200 Subjects release \cite{vanessen2013hcp,smith2013rsfmri}, publicly available via Zenodo (DOI: 10.5281/zenodo.14508470) \cite{roffet2024}. The data consist of $T = 1200$ whole-brain volumes (time samples) per subject, acquired at a repetition time of TR $= 0.72$\,s, giving approximately $14.4$\,min of resting-state data per subject. We use the first resting-state session (REST1) for $N_s=10$ subjects. Whole-brain signals are parcellated into 80 regions of interest using the DBS80 atlas: a Desikan--Killiany-derived cortical parcellation \cite{desikan2006automated} with 31 parcels per hemisphere (62 cortical regions), plus 18 subcortical regions (thalamus, caudate, putamen, globus pallidus externus and internus, subthalamic nucleus, nucleus accumbens, amygdala, and hippocampus, bilaterally). Relative to the full 34-parcel Desikan--Killiany atlas, three small parcels per hemisphere are omitted: the banks of the superior temporal sulcus, the frontal pole, and the temporal pole. The seven Yeo cortical networks of Section~\ref{si:hcp_network} partition the 62 cortical parcels. Regional time series are z-scored to zero mean and unit variance prior to model fitting.

\subsection{VAR(1) model fitting}
For each subject $s$, let $X\in\mathbb{R}^{80\times 1200}$ denote the z-scored regional time series. We define the lagged and led matrices $X_{\text{lag}} = X_{:,1:T-1}\in\mathbb{R}^{80\times(T-1)}$ and $X_{\text{led}} = X_{:,2:T}\in\mathbb{R}^{80\times(T-1)}$ and fit a VAR(1) model
\begin{equation}
\mathbf{x}_{k+1} = \mathbf{A}\mathbf{x}_k + \boldsymbol{\epsilon}_k
\end{equation}
via ordinary least squares:
\begin{equation}
\mathbf{A} = X_{\text{led}}\,X_{\text{lag}}^T\!\left(X_{\text{lag}}\,X_{\text{lag}}^T\right)^{-1}.
\end{equation}
The empirical state and noise covariances are
\begin{equation}
\Sigma = \frac{X_{\text{lag}}\,X_{\text{lag}}^T}{T-2}, \qquad \Sigma_\epsilon = \frac{E\,E^T}{T-2},
\end{equation}
where $E = X_{\text{led}} - \mathbf{A}X_{\text{lag}}$ is the residual matrix. The normalization $T-2$ is the unbiased-covariance convention for the $T-1$ lagged snapshot pairs. The specific choice is immaterial for the results because every reported statistic is a ratio in which this common factor cancels. The spectral radius of $\mathbf{A}$ is verified to be less than one for all subjects, confirming VAR(1) stability.

\subsection{Network assignments}
\label{si:hcp_network}
The 80 regions are assigned to 8 functional groups using the Yeo 7-network cortical parcellation \cite{yeo2011organization} plus a subcortical group: Visual (12 regions), Somatomotor (10), Dorsal Attention (2), Ventral Attention (10), Limbic (8), Frontoparietal (4), Default Mode (16), and Subcortical (18). These define the index sets $\{I_g\}_{g=1}^{8}$ used in the block-level causal statistics.

\subsection{Block-level causal statistics}
\label{si:hcp_block}
For each ordered network pair $(s, r)$ with $s\neq r$, the block propagator $A_{rs} = \mathbf{A}[r, s]\in\mathbb{R}^{|r|\times|s|}$ is extracted from the propagator $\mathbf{A}$ of the full fitted model. The LOCA metric and three-partition Granger statistic are computed per subject as:
\begin{align}
\label{eq:SI_block1}
F^{op}_{s\to r} &= \frac{\mathrm{tr}\!\left(A_{rs}\,\Sigma_{ss}\,A_{rs}^T\right)}{\mathrm{tr}\!\left(\Sigma_{\epsilon,rr}\right)}, \\
\label{eq:SI_block2}
F^{GC}_{s\to r|o} &= \frac{\mathrm{tr}\!\left(A_{rs}\,\Sigma_{s|(r,o)}\,A_{rs}^T\right)}{\mathrm{tr}\!\left(\Sigma_{\epsilon,rr}\right)},
\end{align}
where $\Sigma_{ss} = \Sigma[s,s]$, $\Sigma_{\epsilon,rr} = \Sigma_\epsilon[r,r]$, and $\Sigma_{s|(r,o)}$ is computed from the full 80-dimensional state covariance via Eq.~\eqref{si:eq:sigma_s_given_ro}, with the conditioning set $\mathbf{x}_o$ comprising all 80 regions outside $r\cup s$. The two-partition statistic $F^{GC}_{s\to r}$ is obtained by setting $\mathbf{x}_o=\emptyset$, i.e.\ replacing $\Sigma_{s|(r,o)}$ with $\Sigma_{s|r}$. All statistics are averaged across subjects, and monotonicity is verified numerically for each subject and for the subject-averaged result. 
Note that these block-level statistics can be influenced by the size of the source block, which affects the total number of pathways over which causality is accumulated (and unlike the size of the receiver block, this effect is not removed by the numerator and denominator of Eqs. \ref{eq:SI_block1}-\ref{eq:SI_block2} having the same scaling with this dimension). This effect could be removed by dividing the causality metric by the number of source channels, though this does not change the ratio between Granger and LOCA-based causality measures (nor does it change the conclusions discussed in the main text regarding the Default Mode $\to$ Frontoparietal being the dominant causality link for this dataset).

\subsection{One-lag log-determinant statistics}
In addition to the trace-based statistics above, the one-lag log-determinant (Gaussian transfer entropy) statistics of Eq.~\eqref{eq:TE_Ndim} of the main text are computed for every ordered network pair and subject. Each uses the one-step residual covariance $A_{rs}\,\Sigma_{\mathrm{src}}\,A_{rs}^T + \Sigma_{\epsilon,rr}$ of Eq.~\eqref{eq:Sigma_tilde_main} of the main text, with $\Sigma_{\mathrm{src}}$ set to the unconditional, two-partition, or three-partition source covariance ($\Sigma_s$, $\Sigma_{s|r}$, or $\Sigma_{s|(r,o)}$), giving $TE^{op}$, $TE_{2p}$, and $TE_{3p}$, reported in bits. Penalties are formed from the subject-mean statistics as $1 - TE/TE^{op}$, the same mean-then-ratio convention used for the trace-based penalties. Across the 56 directed pairs the penalties average $69.7\%$ (three-partition) and $34.8\%$ (two-partition), close to the trace-based values of $73.0\%$ and $39.4\%$, and the ordering $TE_{3p} \leq TE_{2p} \leq TE^{op}$ holds with zero violations. These are one-lag statistics: by the proposition of Section~\ref{si:fullhistory}, they upper-bound the corresponding conventional full-history Geweke measures, so the reported penalties are lower bounds on the attenuation of conventional multivariate Granger causality. These bounds are exact for the fitted VAR(1) model, but should be regarded as approximate for the underlying BOLD process.

\subsection{VAR diagnostics and subject-level robustness}
\label{si:fmri_diagnostics}
Three diagnostics assess the adequacy of the VAR(1) specification and the robustness of the reported penalties.

\textit{Lag-order selection.} For each subject, VAR($p$) models with $p = 1,\ldots,5$ are fit by stacked least squares and compared by the multivariate Bayesian information criterion $\mathrm{BIC}(p) = \ln\det\Sigma_\epsilon^{(p)} + p\,n^2 \ln(T_{\mathrm{eff}})/T_{\mathrm{eff}}$, where each additional lag adds $n^2 = 6400$ parameters against $T_{\mathrm{eff}} \approx 1200$ samples. BIC selects $p = 1$ for all ten subjects, with a decisive margin (mean BIC $-56.8$ at $p=1$ versus $-26.0$ at $p=2$), so VAR(1) is the BIC-preferred model among the candidate lag orders considered, rather than an arbitrarily imposed lag choice.%

\textit{Residual whiteness.} If the VAR(1) model captured the dynamics completely, its residuals would be white noise, with no correlation across time. We check how close the fitted residuals come to this ideal in three ways. First, the Ljung--Box test \cite{ljung1978measure}, a standard test for correlation between a series and its own past, is applied to each of the 80 residual series per subject ($L = 20$ lags, $\alpha = 0.01$). Between $49\%$ and $80\%$ of the series pass per subject (mean $62\%$). The test is approximate here because the residuals come from a fitted multivariate model, so we use it as a screen rather than a strict pass/fail criterion. Second, we ask how large the leftover correlation is. The correlation between each residual series and itself one time step earlier has mean absolute value $0.019$--$0.084$ across subjects (mean $0.038$). For comparison, even exactly white residuals would show apparent correlations of about $1/\sqrt{T} \approx 0.029$ from finite-sample estimation alone. The leftover correlation is thus barely above the level that pure estimation noise produces, and is consistent with known physiological and hemodynamic contributions to BOLD signals \cite{kruger2001physiological}. Third, we repeat the size check across channels. Averaging the correlations between all pairs of residual channels at time lags $\ell = 1,\ldots,20$ (the normalized Frobenius norm $\|R_\ell\|_F/n$) gives $0.030$--$0.035$ per subject, with per-subject maxima of $0.034$--$0.065$, again near the finite-sample level. In summary, the residuals are close to but not exactly white. We therefore interpret the causality quantifiers as properties of the fitted one-lag representation rather than as evidence that BOLD dynamics are exactly VAR(1), consistent with the scoping in the main text.

\textit{Subject-level robustness.} The group penalties in the main text pool information across subjects, so a natural question is whether the suppression holds for each subject individually. To check, we compute a single penalty per subject: average the 56 directed-pair statistics within that subject, then form the penalty ratio. Across the 10 subjects this gives $\mathcal{P}_{2p} = 44.5 \pm 6.1\%$ and $\mathcal{P}_{3p} = 76.9 \pm 5.0\%$ (mean $\pm$ SD). The $t$-based $95\%$ confidence intervals are $[40.2, 48.9]\%$ and $[73.3, 80.5]\%$ ($n - 1 = 9$ degrees of freedom for the $n = 10$ subjects). On the log-determinant scale the values are $40.4 \pm 5.4\%$ and $71.4 \pm 4.6\%$, with intervals $[36.5, 44.3]\%$ and $[68.1, 74.7]\%$. The suppression is therefore present in every subject. These values differ modestly from the group values in the main text ($39.4\%$ and $73.0\%$) because the two computations order the same steps differently. The main-text values average each network pair's statistics across subjects first and then form the penalty. The subject-level values form the penalty within each subject first and then average. Averaging and ratio formation do not commute, so the two orderings give slightly different numbers.

\textit{Cohort size.} The ten subjects constitute the complete publicly redistributable release of \cite{roffet2024}, permitting full reproduction of the analysis from openly available data. The fMRI example is intended as a reproducible demonstration of the correlation-penalty mechanism rather than a population-level estimate of human functional connectivity. The confidence intervals above quantify variability within this cohort.

\subsection{Finite-horizon extension}
\begin{figure}[H]
\centering
\includegraphics[width=\linewidth]{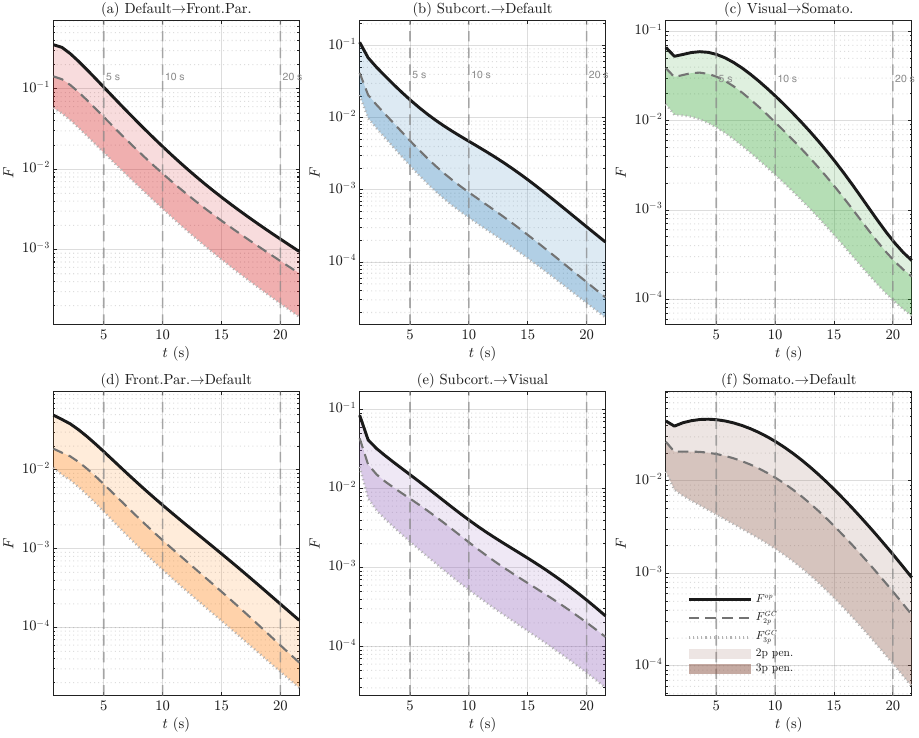}
\caption{Finite-horizon LOCA and Granger causality for six selected HCP functional network pairs. Each panel shows $F^{op}(t)$, $F^{GC}_{2p}(t)$, and $F^{GC}_{3p}(t)$ as functions of horizon time $t=h\cdot\mathrm{TR}$ ($h=1,\ldots,30$, TR $=0.72$\,s), computed from Eq.~\eqref{eq:Fop_group} of the main text with the one-step propagator replaced by $\mathbf{A}^h$. The normalization denominator is fixed at the one-step noise energy $\mathrm{tr}(\Sigma_{\epsilon,rr})$ for all horizons, enabling direct comparison across $h$. The curves are therefore finite-horizon extensions of the one-lag operator-weighted statistics, not conventional $h$-step prediction-error Granger statistics. Shaded bands show the two-partition penalty (light, between $F^{op}$ and $F^{GC}_{2p}$) and the additional three-partition penalty (darker, between $F^{GC}_{2p}$ and $F^{GC}_{3p}$). All curves are subject-averaged. Vertical dashed lines mark $t = 5$, $10$, and $20$\,s. Among the pairs shown, Default Mode $\to$ Frontoparietal (panel a) has the largest $F^{op}$ at short horizons ($t \lesssim 8$\,s) and Somatomotor $\to$ Default Mode (panel f) over most of the longer horizons shown. The three-partition penalty exceeds $0.7$ for every pair at every horizon, and the ordering $F^{GC}_{3p} \le F^{GC}_{2p} \le F^{op}$ holds at every horizon with zero violations.}
\label{fig:si_horizon}
\end{figure}

The block-level statistics above use a single lag. To examine how the suppression evolves over longer horizons, we replace the one-step propagator by $\mathbf{A}^h$ in Eq.~\eqref{eq:Fop_group} of the main text and recompute all three statistics for $h = 1,\ldots,30$ samples ($t = h\cdot\mathrm{TR}$, up to $21.6$\,s). Figure~\ref{fig:si_horizon} shows the result for six representative network pairs. The suppression is not a single-lag artifact. The three-partition penalty exceeds $0.7$ for every pair at every horizon, and the ordering of the three statistics holds at every horizon with zero violations.

\section{Finite-amplitude acoustic Burgers propagation: an exact nonlinear test}
\label{si:acoustic}
This section details the exact nonlinear acoustic test of Fig.~\ref{fig:acoustic} of the main text: the problem setup, the Cole--Hopf solution ensemble, operator identification and noise debiasing, and the amplitude and noise sweeps.

This section provides the full formulation and results for the nonlinear acoustic example summarized in the main text. The objective is to examine how the single-lag causality quantifiers behave on data generated by a continuum nonlinear system rather than an abstract VAR, and to quantify the effect of additive measurement noise on the identified quantities.

\subsection{Problem formulation}
\label{si:acoustic_formulation}
A finite-amplitude plane acoustic wave propagates at the amplitude-dependent speed $c = c_0 + \beta u'$, where $u'$ is the acoustic particle velocity, $c_0 = \sqrt{\gamma R T_0}$ is the small-signal sound speed, and $\beta = (\gamma+1)/2 = 1 + B/2A$ is the coefficient of nonlinearity \cite{beyer1960parameter}, with $B/2A$ the second-order term of the isentropic equation of state. The two contributions to $\beta$ are the local compression/expansion of the medium and the convection of the wave by its own particle velocity. Including thermoviscous absorption through the sound diffusivity $\delta$, and passing to the frame $\xi = x - c_0 t$ that removes the mean drift, the particle velocity obeys the Burgers equation of nonlinear acoustics \cite{burgers1948mathematical,hamilton1998nonlinear},
\begin{equation}\label{si:eq:burgers_acoustic}
\frac{\partial u}{\partial t} + \beta\,u\,\frac{\partial u}{\partial \xi} = \delta\,\frac{\partial^2 u}{\partial \xi^2}.
\end{equation}
Nonlinear steepening competes with dissipation to set a finite shock-formation time. For an initial disturbance $u_0(\xi)$, the inviscid estimate is
\begin{equation}\label{si:eq:tsh}
t_{sh} = \frac{1}{\beta\,\max_\xi\left(-\partial_\xi u_0\right)},
\end{equation}
which uses the peak compressive (most-negative) slope of the source because that slope governs characteristic crossing. Equation~\eqref{si:eq:tsh} is the temporal analog of the plane-wave shock-formation distance $x_{sh} = \frac{\gamma}{\pi(\gamma+1)}(p_0/p')\,\lambda$ (equivalently $\rho_0 c_0^3/(\beta\omega p')$, with $p'$ the pressure amplitude) \cite{blackstock1966connection}. Two dimensionless parameters locate the operating point: the nonlinearity parameter $\beta M_a$, where $M_a = \max|u_0|/c_0$ is the peak acoustic Mach number, and the ratio $\Pi \equiv T/t_{sh}$ of the propagation horizon $T$ to $t_{sh}$, which spans the range from the linear-acoustics limit ($\Pi \ll 1$) to shocked propagation ($\Pi \gtrsim 1$). The sweep below spans $\beta M_a$ from $3\times 10^{-4}$ up to $\approx 0.51$. The upper end of this range is a fully nonlinear Burgers continuation of the finite-amplitude acoustic model, used here as an exact-solution stress test rather than a quantitatively small-amplitude acoustic prediction. The small-signal interpretation of $\beta$ and $M_a$ applies strictly only near the linear plateau.

\subsection{Exact Cole--Hopf solution and stochastic ensemble}
\label{si:acoustic_colehopf}
The value of this test case is that the nonlinear dynamics can be solved exactly, so no numerical-integration error enters the ensemble. Equation~\eqref{si:eq:burgers_acoustic} is solved exactly by the Cole--Hopf transformation \cite{cole1951quasilinear,hopf1950partial}, so every realization comes from the exact solution rather than a time-stepped approximation. The only remaining numerical errors are spectral truncation on the finite periodic grid and interpolation to the probe locations, and both are checked to be below the statistical resolution of the ensemble. Writing $v = \beta u$,
\begin{equation}\label{si:eq:colehopf}
v(\xi,t) = -2\delta\,\frac{\partial_\xi\theta}{\theta}, \qquad \frac{\partial\theta}{\partial t} = \delta\,\frac{\partial^2\theta}{\partial\xi^2}, \qquad \theta(\xi,0) = \exp\!\left[-\frac{1}{2\delta}\int_0^{\xi} v_0(\xi')\,\mathrm{d}\xi'\right],
\end{equation}
so the nonlinear problem reduces to the linear heat equation for $\theta$, which is evolved exactly by the heat kernel (implemented spectrally). The source is a windowed single-cycle disturbance $u_0(\xi) = a\,\sin[2\pi(\xi-\xi_0)/w]\,\exp\!\left[-\left((\xi-\xi_0)/w\right)^2\right]$. This profile is odd about $\xi_0$, so its net integral vanishes by symmetry (to numerical precision on the periodic grid, since the Gaussian window is fully resolved). As a result, $\theta(\xi,0)$ returns to a constant at the domain ends and remains periodic and bounded.

\begin{figure}[!htbp]
\centering
\includegraphics[width=0.55\linewidth]{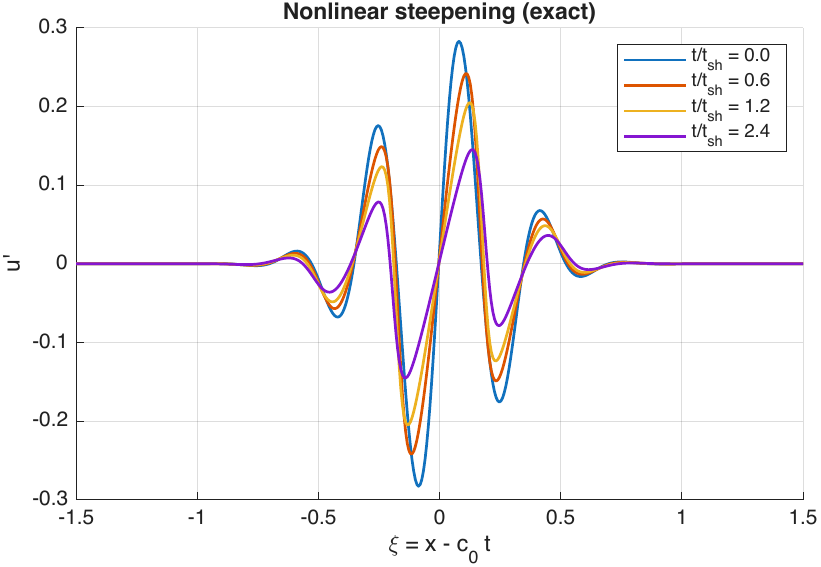}
\caption{Exact Cole--Hopf waveform in the moving frame at successive $t/t_{sh}$: nonlinear steepening toward an $N$-wave and the associated earlier arrival relative to $c_0 t$.}
\label{fig:si_acoustic_steepening}
\end{figure}

We place all of the stochastic content in the source and not in the PDE, for a practical reason. If Eq.~\eqref{si:eq:burgers_acoustic} were forced with continuous random noise, that forcing would appear in the transformed heat equation as a random coefficient multiplying $\theta$ rather than as a simple additive input, and the exact Cole--Hopf solution would be lost. Instead, each ensemble member is an exact deterministic realization with a randomized upstream amplitude $a = \bar{a}(1+\sigma_a\eta_a)$ and source location $\xi_0 = \sigma_\xi\eta_\xi$, where $\eta_a,\eta_\xi \sim \mathcal{N}(0,1)$. All identified quantities come from a pooled one-step least-squares regression over this ensemble of transient solutions: covariances $\Sigma$, correlations $\rho$, and the source variance are pooled ensemble statistics. No stationarity assumption is required for the causality identity below, which follows from least-squares projection geometry alone. A chain of $N=8$ probes at spacing $\Delta x_p$ is sampled stroboscopically at the inter-probe transit interval $\Delta t = \Delta x_p/c_0$, so one time step advances the disturbance by one probe. The single lag is deliberately fixed to the linear transit time. Finite-amplitude departures, namely earlier arrival and waveform steepening, are therefore not absorbed into a re-timed lag. Instead, they appear as changes in the identified operator and its residual, which is exactly the effect the sweep is designed to isolate. Figure~\ref{fig:si_acoustic_steepening} shows the exact waveform steepening and earlier arrival as $t/t_{sh}$ increases.

\subsection{Identified operator and measurement-noise debiasing}
\label{si:acoustic_debiasing}
The single-lag operator on the full observed probe vector is estimated by least squares (i.e., exact DMD on the probe states) \cite{schmid2010dynamic,tu2014dynamic},
\begin{equation}\label{si:eq:dmd}
A_{\mathrm{DMD}} = \tilde{X}_1\tilde{X}_0^\top\!\left(\tilde{X}_0\tilde{X}_0^\top\right)^{-1}, \qquad \Sigma_\epsilon = \mathrm{cov}\!\left(\tilde{X}_1 - A_{\mathrm{DMD}}\tilde{X}_0\right), \qquad \Sigma = \mathrm{cov}\!\left(\tilde{X}_0\right),
\end{equation}
where $\tilde{X}_0,\tilde{X}_1$ are the lagged and led probe-snapshot matrices, mean-centered over the ensemble. Covariances use the $n_s-1$ normalization, with $n_s$ the number of snapshot pairs. When the probes carry additive measurement noise of variance $\sigma_n^2$, the identification is biased because the noise now sits on the regressors (the input snapshots), not just on the output, known as an errors-in-variables problem. The effect is familiar from frequency-response estimation: uncorrelated noise on the input inflates the measured input autospectrum and biases the $H_1$ estimator low. The same thing happens here in the time domain. The noise inflates the measured state covariance, so the least-squares estimate suffers the attenuation bias $A_{\mathrm{DMD}} \to A\,\Sigma(\Sigma+\sigma_n^2 I)^{-1}$, and $\Sigma_\epsilon$ is inflated because the fit cannot explain the noise. Under this additive sensor-noise model, we correct the attenuation using the noise-covariance subtraction of debiased DMD \cite{dawson2016dmdnoise}, which removes the known noise contribution from each estimated quantity before it is used. Specifically, the regressor Gram $\tilde{X}_0\tilde{X}_0^\top$ is replaced by $\tilde{X}_0\tilde{X}_0^\top - \sigma_n^2(n_s-1)I$, which is the time-domain analog of subtracting the noise autospectrum from the input autospectrum before forming a frequency-response estimate. Similarly, $\Sigma_\epsilon$ is de-inflated by $\sigma_n^2(I + A_{\mathrm{debiased}}A_{\mathrm{debiased}}^\top)$ (using the debiased operator itself in this term), and $\Sigma$ is de-inflated by $\sigma_n^2 I$. This correction targets the measurement-noise attenuation alone and leaves the physical (nonlinear) residual intact.\footnote{We do not use a total-least-squares debiaser (total DMD) \cite{hemati2017tls} here because the one-step linear surrogate has appreciable nonlinear model discrepancy in addition to measurement noise, so a generic errors-in-variables treatment does not cleanly distinguish these contributions. In our numerical tests it over-corrects the operator, whereas the noise-covariance subtraction recovers the clean operator to $O(10^{-3})$. On an exact linear surrogate the two debiasers coincide.} Separately, and purely as a numerical safeguard, the debiased covariances are symmetrized and projected onto the positive semidefinite cone (eigenvalues clipped at a relative floor of $10^{-12}$). This guarantees that the conditional variances in Section~\ref{si:acoustic_quantifiers} remain well posed (no negative variances) when subtraction leaves a marginally indefinite matrix at high noise. The projection stabilizes the conditional-variance computation and is not part of the correction itself. Across the sweep, it perturbs the covariances by at most $\approx 8.2\times 10^{-8}$ and does not affect any reported quantity.

\subsection{Linearized reference operator and nonlinearity residual}
\label{si:acoustic_linref}
\begin{figure}[!htbp]
\centering
\includegraphics[width=\linewidth]{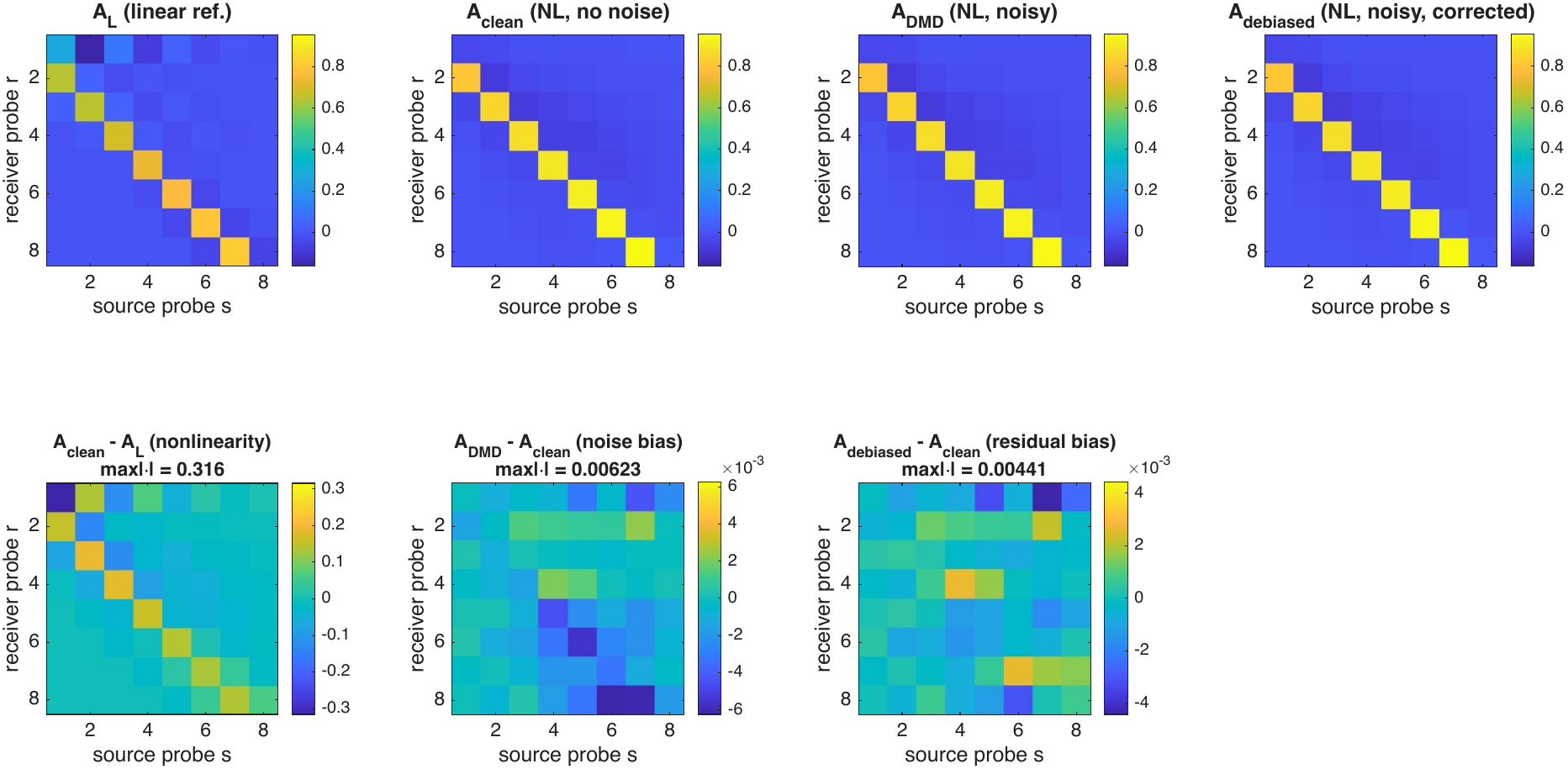}
\caption{Identified operators on the reduced probe space at $\bar{a}=0.22$, $5\%$ rms sensor noise: linear reference $A_L$, clean nonlinear $A_{\mathrm{clean}}$, noisy exact-DMD estimate $A_{\mathrm{DMD}}$, debiased $A_{\mathrm{debiased}}$, and the three difference maps isolating nonlinearity, noise bias, and residual bias. The three difference maps use separate symmetric color scales, with the maximum absolute entry reported in each title.}
\label{fig:si_acoustic_operators}
\end{figure}

To measure how nonlinear the identified dynamics have become, we need a linear reference to compare against. Linearizing Eq.~\eqref{si:eq:burgers_acoustic} about the quiescent state gives pure diffusion in the moving frame, $\partial_t u = \delta\,\partial^2_\xi u$, i.e.\ linear advection--diffusion $\partial_t u + c_0\partial_x u = \delta\,\partial^2_x u$ in the laboratory frame. Its exact solution (heat-kernel evolution of the source, translating at $c_0$) is sampled on the identical probe array and ensemble design, and the linearized operator $A_L$ comes from the same least-squares fit applied to these linear data. $A_L$ is therefore defined on exactly the same reduced observation space as $A_{\mathrm{DMD}}$, so the two are directly comparable. Numerically, $A_L$ shows the expected near-shift structure: dominant subdiagonal transport coupling (mean subdiagonal $0.731$) and a small diagonal (mean $0.033$). To isolate the finite-amplitude dependence of the identified operator, we use the normalized nonlinearity residual
\begin{equation}\label{si:eq:R}
\mathcal{R}(\bar{a}) = \frac{\left\|A_{\mathrm{clean}} - A_L\right\|_F}{\left\|A_L\right\|_F},
\end{equation}
where $A_{\mathrm{clean}}$ is the noise-free least-squares operator at amplitude $\bar{a}$. $\mathcal{R}$ measures the amplitude dependence of the identified reduced operator, not the full PDE nonlinearity. As Table~\ref{tab:si_acoustic_sweep} reports, $\mathcal{R}$ grows monotonically from $0.013$ to $0.586$ across the amplitude sweep and is the explicit fingerprint of the nonlinearity in the identified model. Figure~\ref{fig:si_acoustic_operators} shows $A_L$, $A_{\mathrm{clean}}$, $A_{\mathrm{DMD}}$ and the debiased operator, together with the difference maps that separate the nonlinearity ($A_{\mathrm{clean}}-A_L$) from the measurement-noise bias ($A_{\mathrm{DMD}}-A_{\mathrm{clean}}$) and the residual bias after debiasing.

\subsection{Causality quantifiers and the correlation penalty}
\label{si:acoustic_quantifiers}
The causality quantifiers for the probe array follow the same construction as in the chain example. For a source probe $s$ and receiver probe $r$, with the remaining probes as the conditioning set $\mathbf{x}_o$, the single-lag three-partition quantifiers are the scalar-receiver case of Eqs.~\eqref{eq:Fop_group} and~\eqref{eq:Granger-N-dim-F} of the main text:
\begin{equation}\label{si:eq:acoustic_quantifiers}
F^{op}_{s\to r} = \frac{A_{rs}^2\,\Sigma_{ss}}{\Sigma_{\epsilon,rr}}, \qquad F^{GC}_{s\to r|o} = \frac{A_{rs}^2\,\Sigma_{s|(r,o)}}{\Sigma_{\epsilon,rr}}, \qquad \mathcal{P}_{s\to r|o} = 1 - \frac{\Sigma_{s|(r,o)}}{\Sigma_{ss}},
\end{equation}
where $\Sigma_{s|(r,o)} = \Sigma_{ss} - \Sigma_{sz}\Sigma_z^{-1}\Sigma_{zs}$ is the conditional (Schur-complement) source variance over the retained predictors $z = (r,o)$, i.e.\ all probes except the source, exactly as in Eq.~\eqref{si:eq:sigma_s_given_ro}. In plain terms, $\Sigma_{s|(r,o)}$ is the part of the source variance that cannot be predicted from the other probes. $\mathcal{P}_{s\to r|o}$ is the three-partition penalty of Eq.~\eqref{eq:penalty_Ndim} of the main text (the multiple $R^2$ of the source on the retained predictors, written $P$ in the figure axes), reducing to the squared correlation $\rho^2$ in the bivariate case. Here $F^{GC}$ is the variance-weighted single-lag Granger-causality statistic used throughout this paper \cite{granger1969}, not the classical log-ratio $\log(\Sigma^{\mathrm{restricted}}_{\epsilon,rr}/\Sigma_{\epsilon,rr})$, and $F^{op}$ is its full-model operator counterpart. These quantifiers satisfy $F^{GC}_{s\to r|o} = F^{op}_{s\to r}(1-\mathcal{P}_{s\to r|o})$ identically, so $\mathcal{P}_{s\to r|o}$ is exactly the fractional correlation penalty by which single-lag Granger causality under-reports the operator pathway strength, consistent with the generative formalism for causality quantifiers of \cite{smirnov2022generative}. Because both quantifiers are built from the unrestricted operator, the same coefficient $A_{rs}$ appears in each. The restricted-model coefficients themselves do shift when the source is omitted (Eq.~\eqref{eq:Atilde_rz} of the main text). After accounting for that shift analytically, the increase in prediction-error variance depends only on the unrestricted $A_{rs}$ and the conditional source variance, so the penalty is non-negative. All reported scalar $F^{op}$, $F^{GC}$, and $\mathcal{P}$ values are for source probe 4 to receiver probe 5 (two interior probes), i.e.\ $s=\{4\}$ and $r=\{5\}$, with the remaining six probes forming the conditioning set.

\subsection{Amplitude sweep}
\label{si:acoustic_sweep}
\begin{figure}[!htbp]
\centering
\includegraphics[width=0.85\linewidth]{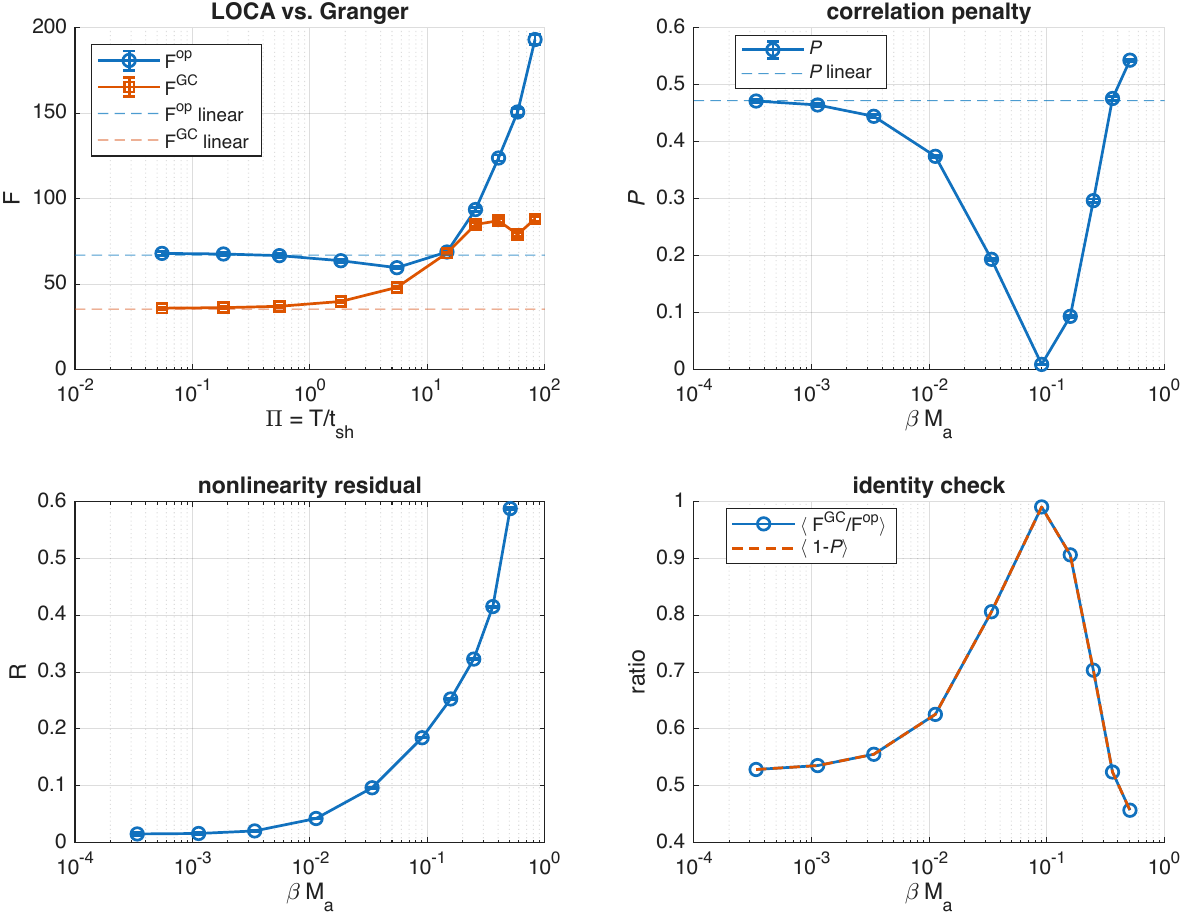}
\caption{Amplitude/shock-proximity sweep: $F^{op}$ and $F^{GC}$ (with their linear asymptotes), the penalty $\mathcal{P}$ against its linear value, the nonlinearity residual $\mathcal{R}$, and the identity check $\langle F^{GC}/F^{op}\rangle = \langle 1-\mathcal{P}\rangle$ (both averaged seed-by-seed). The first panel uses $\Pi = T/t_{sh}$ and the remaining panels use $\beta M_a$. Both are monotonic in $\bar{a}$, so the horizontal ordering of points is identical across panels.}
\label{fig:si_acoustic_sweep}
\end{figure}

Table~\ref{tab:si_acoustic_sweep} and Fig.~\ref{fig:si_acoustic_sweep} report the sweep in $\bar{a}$ (equivalently $\Pi = T/t_{sh}$), where each entry is averaged over 8 seeds with the standard error shown. The identity $F^{GC} = F^{op}(1-\mathcal{P})$ holds to machine precision at every amplitude. This is a consequence of least-squares projection geometry: once the operator, covariances, and restricted regression are computed from the same pooled data, the equality is algebraically guaranteed, requiring neither stationarity nor Gaussianity nor a linear generative model. A direct restricted-model refit (regressing the receiver on all lagged probes except the source) reproduces the formula-based $F^{GC}$ to $\sim 10^{-14}$ relative error, confirming the implementation. The nonlinear sweep therefore does not independently validate the identity but probes how the quantities entering it evolve with the dynamics. We also verify the discrete Lyapunov relation for the fitted quantities: $\|\Sigma - A\Sigma A^\top - \Sigma_\epsilon\|/\|\Sigma\| = 2.62\times 10^{-6}$. Least-squares orthogonality guarantees $\Sigma_{\text{led}} = A\Sigma_{\text{lag}}A^\top + \Sigma_\epsilon$ exactly, where $\Sigma_{\text{lag}}$ and $\Sigma_{\text{led}}$ are the pooled covariances of the snapshots before and after one step. The Lyapunov residual therefore measures only the difference between the two. Its small value confirms that the pooled covariance is essentially unchanged over one step, which is the consistency assumed when a single $\Sigma$ is used in the source statistics. It does not test stationarity of the individual transient realizations. The sweep separates two distinct behaviors. First, the nonlinearity residual $\mathcal{R}$ increases monotonically from $0.013$ to $0.586$. In contrast, the correlation penalty $\mathcal{P}$ is non-monotonic. At $\Pi \ll 1$, the penalty is set by the linear near-shift correlation of neighboring probes ($\mathcal{P} \approx 0.47$, the visible linear plateau). It collapses to $\mathcal{P} \approx 0.01$ near $\Pi \approx 14.7$ and then increases to $\mathcal{P} \approx 0.54$ as the $N$-wave develops. The non-monotonicity is therefore a property of the correlation structure, not the operator identification.

\begin{table}[!htbp]
\centering
\caption{Amplitude sweep (nominal $\bar{a}$; 8 seeds; three significant figures). Standard errors are shown for $F^{op}$, $F^{GC}$, and $\mathcal{P}$. The $\mathcal{R}$ standard errors are $\leq 0.004$ and omitted. $\beta M_a$ is the nonlinearity parameter ($M_a = \max|u_0|/c_0$ the peak acoustic Mach number) and $\Pi = T/t_{sh}$ the shock-proximity ratio.}
\label{tab:si_acoustic_sweep}
\begin{tabular}{ccccccc}
\toprule
$\bar{a}$ & $\beta M_a$ & $\Pi = T/t_{sh}$ & $F^{op}$ & $F^{GC}$ & $\mathcal{P}$ & $\mathcal{R}$ \\
\midrule
0.0003 & $3.39\times 10^{-4}$ & 0.0551 & $68.2 \pm 1.7$ & $36.0 \pm 0.9$ & $0.472 \pm 0.003$ & 0.013 \\
0.0010 & $1.13\times 10^{-3}$ & 0.184 & $67.8 \pm 1.6$ & $36.2 \pm 0.9$ & $0.465 \pm 0.003$ & 0.013 \\
0.0030 & $3.39\times 10^{-3}$ & 0.551 & $66.8 \pm 1.6$ & $37.0 \pm 0.9$ & $0.446 \pm 0.003$ & 0.019 \\
0.0100 & $1.13\times 10^{-2}$ & 1.84 & $63.7 \pm 1.4$ & $39.8 \pm 0.8$ & $0.376 \pm 0.003$ & 0.042 \\
0.0300 & $3.39\times 10^{-2}$ & 5.51 & $59.4 \pm 1.0$ & $47.8 \pm 0.8$ & $0.195 \pm 0.003$ & 0.097 \\
0.0800 & 0.090 & 14.7 & $68.3 \pm 0.8$ & $67.6 \pm 0.8$ & $0.009 \pm 0.000$ & 0.186 \\
0.1400 & 0.158 & 25.7 & $92.8 \pm 1.4$ & $84.3 \pm 1.2$ & $0.092 \pm 0.003$ & 0.255 \\
0.2200 & 0.248 & 40.4 & $124 \pm 2.4$ & $87.2 \pm 1.6$ & $0.295 \pm 0.004$ & 0.325 \\
0.3200 & 0.361 & 58.7 & $153 \pm 3.3$ & $80.3 \pm 1.6$ & $0.474 \pm 0.004$ & 0.417 \\
0.4500 & 0.508 & 82.6 & $196 \pm 4.5$ & $89.8 \pm 1.9$ & $0.543 \pm 0.003$ & 0.586 \\
\bottomrule
\end{tabular}
\end{table}

\subsection{Origin of the non-monotonic penalty}
\label{si:acoustic_penalty_origin}
\begin{figure}[!htbp]
\centering
\includegraphics[width=0.55\linewidth]{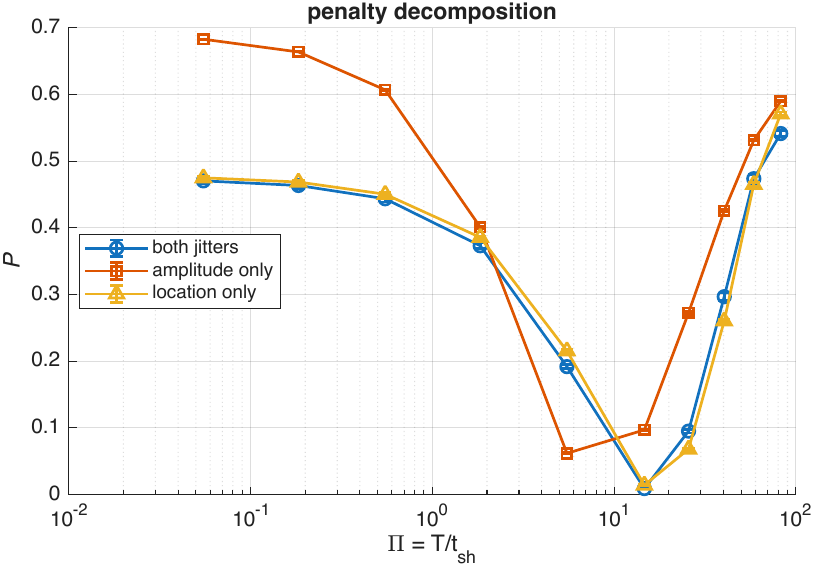}
\caption{Penalty $\mathcal{P}$ versus shock proximity for the combined ensemble and the amplitude-only and location-only ablations, showing the competing correlation mechanisms behind the non-monotonic penalty.}
\label{fig:si_acoustic_ablation}
\end{figure}

\begin{figure}[!htbp]
\centering
\includegraphics[width=0.9\linewidth]{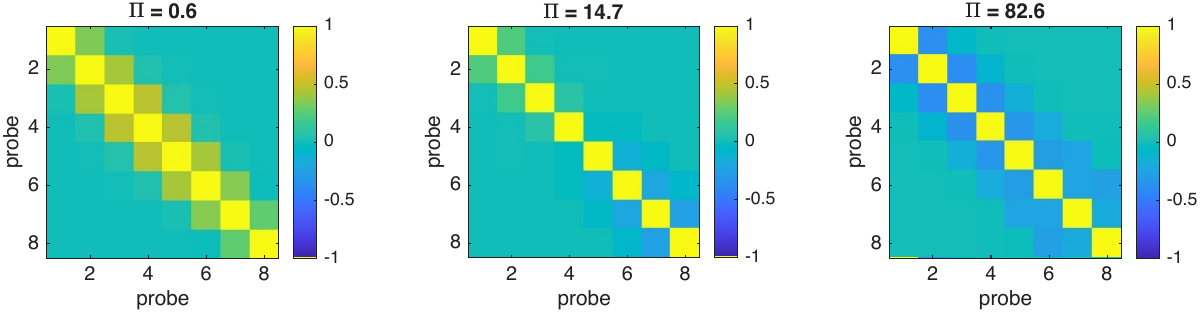}
\caption{Probe correlation matrices at low, intermediate, and high shock proximity: strong near-neighbor correlation on the linear plateau, collapse of shared structure near the penalty minimum, and recovery of structured correlation once the mature $N$-wave forms.}
\label{fig:si_acoustic_corr}
\end{figure}

The dip in $\mathcal{P}$ can be understood by asking what each of the two source randomizations does to the probe signals. Amplitude jitter scales the entire waveform up or down, so all probes move together. This shared variation is exactly what correlates the probes, and at small $\Pi$ it produces the largest penalty ($\mathcal{P} \approx 0.68$ for the amplitude-only ensemble in Fig.~\ref{fig:si_acoustic_ablation}). Location jitter instead shifts the waveform slightly upstream or downstream by a random amount of size $\sigma_\xi$. Whether such a shift matters depends on how steep the waveform is where the probes sample it: shifting a smooth, gentle waveform by $\sigma_\xi$ barely changes any probe reading, but shifting a waveform with a steep front by the same amount changes the reading at that front dramatically. Two length scales, both compared to $\sigma_\xi$, make this precise. The overall width of the disturbance does not shrink. Its autocorrelation half-length $\ell$ actually grows from $0.70\,\sigma_\xi$ to $1.90\,\sigma_\xi$ across the sweep as the wave spreads. What shrinks is the thickness of the shock front, estimated by the nominal Taylor-shock scale $\ell_{TS} = 4\delta/(\beta\,\Delta u_0)$, where $\Delta u_0 = \max_\xi u_0 - \min_\xi u_0$ is the source velocity jump (a source-based estimate, not a measurement of the evolved front). As the amplitude increases, $\ell_{TS}/\sigma_\xi$ falls from $\approx 158$, where the front is far wider than the shift and the shift is invisible, through unity, to $\approx 0.11$, where the front is far thinner than the shift and a small shift moves the entire front past a probe. Near $\Pi \approx 14.7$ this effect is at its worst: the front is razor thin, so small random source shifts swing the probe readings wildly and nearly independently, the probes decorrelate, and $\mathcal{P}$ collapses to its floor ($\approx 0.01$). The ablation confirms that the location jitter is responsible: the location-only and combined curves collapse together at the floor, while the amplitude-only curve stays higher ($\approx 0.09$). At still larger $\Pi$, the mature $N$-wave replaces the isolated thin front with a long, gently sloping sawtooth ramp. Most probes then sample the smooth ramp, their readings move together again, and $\mathcal{P}$ recovers. Figure~\ref{fig:si_acoustic_corr} shows the corresponding probe correlation matrices at low, intermediate, and high shock proximity ($\Pi = T/t_{sh}$ well below, near, and beyond unity), giving the spatial explanation that the scalar $\mathcal{P}$ curve alone cannot.

\subsection{Measurement-noise sensitivity}
\label{si:acoustic_noise}
\begin{figure}[!htbp]
\centering
\includegraphics[width=0.85\linewidth]{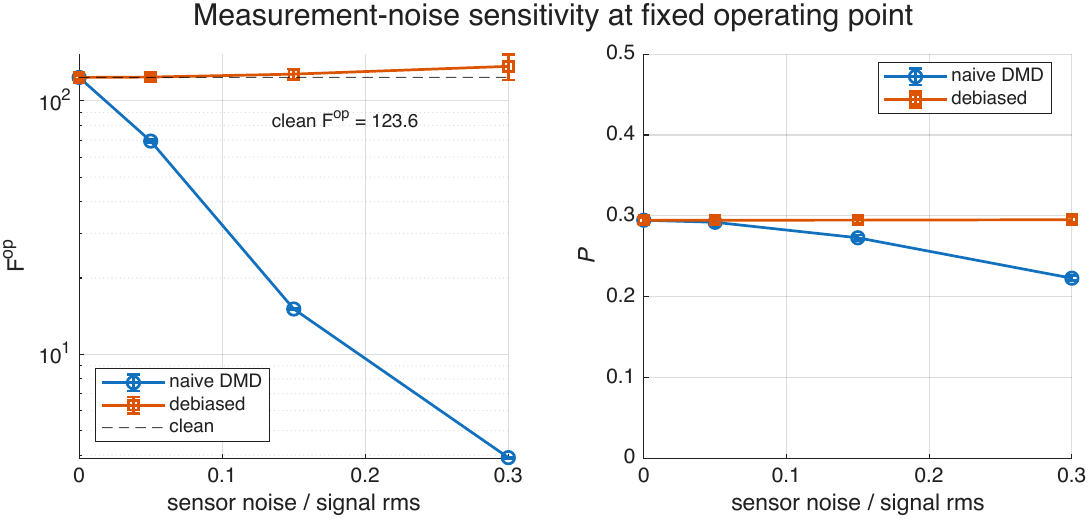}
\caption{Measurement-noise sensitivity at $\bar{a}=0.22$ (clean $F^{op} = 123.6$, annotated on the left panel): exact DMD $F^{op}$ collapses with sensor noise while the debiased estimate recovers the clean value through moderate noise and becomes increasingly variable at the highest noise level tested. The penalty $\mathcal{P}$ is comparatively robust for both.}
\label{fig:si_acoustic_noise}
\end{figure}

Table~\ref{tab:si_acoustic_noise} and Fig.~\ref{fig:si_acoustic_noise} fix the operating point at $\bar{a} = 0.22$ (clean $F^{op} = 123.6 \pm 2.4$) and sweep the sensor noise. Exact DMD attenuates $F^{op}$ severely ($123.6 \to 3.9$ as the noise reaches $30\%$ of signal rms), consistent with the errors-in-variables attenuation described in Section~\ref{si:acoustic_debiasing}. Even a small Frobenius-norm operator bias produces a large $F^{op}$ bias because sensor noise inflates the residual variance $\Sigma_{\epsilon,rr}$ in the denominator of $F^{op}$ far more than it perturbs the structural coefficient $A_{rs}$ in the numerator. For example, at $5\%$ noise the operator changes by only $\|A_{\mathrm{DMD}}-A_{\mathrm{clean}}\|_F/\|A_{\mathrm{clean}}\|_F \approx 0.007$, yet $F^{op}$ falls from $123.6$ to $69.4$. Noise-covariance subtraction recovers the operator metric through moderate sensor noise ($5\%$ and $15\%$). At $30\%$ of signal rms the corrected estimate exhibits substantially greater variance and sensitivity to estimator details (the standard error grows from $\pm 2.4$ clean to $\pm 16.0$, with mean $136.8$ versus clean $123.6$), although it remains far less attenuated than exact DMD. Conditioning diagnostics rule out matrix ill-conditioning as the cause: the reciprocal condition number of the noise-subtracted regressor Gram is essentially unchanged across the sweep (worst case $0.062$ at $30\%$ versus $0.065$ clean, with smallest eigenvalue $\approx 0.022$ throughout). The behavior is consistent with increasing cancellation between large terms in the residual-covariance subtraction: the correction is exact in expectation, but the relative sampling error of the subtracted result grows with the noise level, and with it the error of $F^{op}$, whose denominator it supplies. The penalty $\mathcal{P}$ is comparatively robust for both estimators. It drifts only from $0.295$ to $0.223$ for exact DMD and remains constant ($0.295$) after debiasing. The ratio-based penalty is therefore far less sensitive to sensor noise than the absolute causality magnitudes.

\begin{table}[!htbp]
\centering
\caption{Measurement-noise sweep at $\bar{a}=0.22$ (clean $F^{op} = 123.6 \pm 2.4$). Exact DMD versus debiased (noise-covariance-subtracted) operator identification. Each $F^{op}$ entry is the mean $\pm$ standard error over the 8 noise-realization seeds. Each $\mathcal{P}$ entry is the mean over the same seeds.}
\label{tab:si_acoustic_noise}
\begin{tabular}{ccccc}
\toprule
noise/rms & $F^{op}$ DMD & $F^{op}$ debiased & $\mathcal{P}$ DMD & $\mathcal{P}$ debiased \\
\midrule
0.00 & $123.6 \pm 2.4$ & $123.6 \pm 2.4$ & 0.295 & 0.295 \\
0.05 & $69.4 \pm 1.0$ & $124.0 \pm 2.8$ & 0.292 & 0.295 \\
0.15 & $15.1 \pm 0.11$ & $127.4 \pm 6.0$ & 0.273 & 0.295 \\
0.30 & $3.9 \pm 0.03$ & $136.8 \pm 16.0$ & 0.223 & 0.295 \\
\bottomrule
\end{tabular}
\end{table}

\bibliographystyle{pnas-new}
\bibliography{Master}

\end{document}